%% file: sxdf_xclus_astroph.tex
\documentclass[useAMS,usenatbib]{mn2e}
\usepackage{float,epsfig,deluxetable}
\usepackage{pstricks,longtable}
\usepackage{times}

\newcommand{\mincir}{\raise
  -2.truept\hbox{\rlap{\hbox{$\sim$}}\raise5.truept \hbox{$<$}\ }}
\newcommand{\magcir}{\raise
  -2.truept\hbox{\rlap{\hbox{$\sim$}}\raise5.truept \hbox{$>$}\ }}

\def\deg      {{\ifmmode^\circ\else$^\circ$\fi}} 

\begin{document}

\title[Clusters in SXDF]{X-ray groups and clusters of
  galaxies in the Subaru-XMM Deep Field}

\author[Finoguenov et al.]{A. Finoguenov$^{1,2}$, M.G. Watson$^3$,
  M. Tanaka$^4$, C. Simpson$^5$, \newauthor 
M. Cirasuolo$^6$,  J.S. Dunlop$^6$,
J.A. Peacock$^6$, D. Farrah$^{7}$, M. Akiyama$^{8}$, Y. Ueda$^{9}$, V.~Smol\v{c}i\'{c}$^{15}$,
\newauthor  
G. Stewart$^3$, S. Rawlings$^{10}$, C. van Breukelen$^{10}$, O. Almaini$^{11}$,
L. Clewley$^{10}$, D.G. Bonfield$^{12}$, \newauthor M.J. Jarvis$^{12}$,
J.M. Barr$^{10}$, S. Foucaud$^{11}$, R.J. McLure$^6$,   K. Sekiguchi$^{13}$, E. Egami$^{14}$\\
{$^1$ Max-Planck-Institut f\"ur extraterrestrische Physik,
             Giessenbachstra\ss e, 85748 Garching, Germany}\\
{$^2$ University of Maryland, Baltimore County, 1000
  Hilltop Circle,  Baltimore, MD 21250, USA}\\
{$^3$ X-ray Astronomy Group, Department of Physics and Astronomy, University
  of Leicester, Leicester LE1 7RH}\\
{$^4$ European Southern Observatory, Karl-Schwarzschild-Str 2, 85748, Garching, Germany}\\
{$^5$ Astrophysics Research Institute, Liverpool John Moores University, Twelve Quays House, Egerton Wharf, Birkenhead CH41 1LD}\\
{$^6$ Scottish Universities Physics Alliance, Institute for Astronomy, University of Edinburgh, Royal Observatory, Edinburgh EH9 3HJ}\\
{$^{7}$ Dept of Physics \& Astronomy, University of Sussex, Falmer, Brighton, UK}\\
{$^{8}$ Astronomical Institute, Tohoku University, Sendai 980-8578, Japan}\\
{$^{9}$ Department of Astronomy, Kyoto University, Kyoto 606-8502, Japan}\\
{$^{10}$ Astrophysics, Department of Physics, Keble Road, Oxford OX1 3RH}\\
{$^{11}$ School of Physics and Astronomy, University of Nottingham,
University Park, Nottingham NG7 2RD} \\
{$^{12}$  Centre for Astrophysics Research, Science \& Technology Research Institute, University of Hertfordshire, Hatfield, AL10 9AB, UK}\\
{$^{13}$ Subaru Telescope, National Astronomical Observatory of Japan,
650 N. A'ohoku Place, Hilo, Hawaii 96720, USA}  \\
{$^{14}$ Steward Observatory, University of Arizona, 933 N. Cherry
  Ave., Tucson, AZ  85721} \\
{$^{15}$ California Institute of Technology, MC 105-24, 1200 East
California Boulevard, Pasadena, CA 91125 }
}

\date{subm. to MNRAS, Sep. 24 2009}

\pagerange{\pageref{firstpage}--\pageref{lastpage}} \pubyear{2009}

\maketitle

\label{firstpage}

\begin{abstract}
  We present the results of a search for galaxy clusters in Subaru-XMM Deep
  Field.  We reach a depth for a total cluster flux in the 0.5--2 keV band
  of $2\times10^{-15}$ ergs cm$^{-2}$ s$^{-1}$ over one of the widest
  XMM-Newton contiguous raster surveys, covering an area of 1.3 square
  degrees. Cluster candidates are identified through a wavelet detection of
  extended X-ray emission. The red sequence technique allows us to identify
  57 cluster candidates.  We report on the progress with the cluster
  spectroscopic follow-up and derive their properties based on the X-ray
  luminosity and cluster scaling relations. In addition, 3 sources are
  identified as X-ray counterparts of radio lobes, and in 3 further sources,
  X-ray counterpart of radio lobes provides a significant fraction of the
  total flux of the source.  In the area covered by NIR data, our
  identification success rate achieves 86\%. We detect a number of radio
  galaxies within our groups and for a luminosity-limited sample of radio
  galaxies we compute halo occupation statistics using a marked cluster mass
  function. We compare the cluster detection statistics in the SXDF with the
  predictions of concordance cosmology and current knowledge of the X-ray
  cluster properties, concluding that a reduction of concordance $\sigma_8$
  value by 5\% is required in order to match the prediction of the model and
  the data. This conclusion still needs verification through the completion
  of cluster follow-up.
\end{abstract}


\section{Introduction}

Extended X-ray emission from groups and clusters of galaxies is an
unambiguous signal of high density, high mass environments (e.g.  Borgani \&
Guzzo 2001, Rosati et al. 2002). The low scatter of X-ray emission around
the mean with respect to the underlying mass of the object and advances in
X-ray surveys, have established X-rays as one of the most reliable tools in
the search for massive halos (e.g. B\"ohringer et al 2002). Deep XMM and
Chandra surveys such as CDFS (Giacconi et al. 2002), CDFN (Bauer et al.
2002), Lockman Hole (Finoguenov et al. 2005), COSMOS (Finoguenov et al.
2007), XMM-LSS (Pacaud et al. 2007), CNOC2 (Finoguenov et al. 2009) show the
potential of efficient group/cluster detection and illustrate their
competitiveness with spectroscopic group surveys. Such data have contributed
directly to studies of galaxy formation (e.g. Tanaka et al.  2008; Giodini
et al. 2009), Large Scale Structure (LSS) and its relation to AGN activity
(Silverman et al.  2009), and have also shown the power of X-ray surveys to
find and study sky densities in excess of 100 groups per square degree
(Bauer et al.  2002).

At high redshifts, deep X-ray surveys offer both the highest sensitivity
towards the cluster mass and are competitive to the best optical surveys for
finding groups. Clusters at different redshifts provide homogeneous samples
of galaxies in a high-density environment, enabling studies of the evolution
of stellar populations (e.g. Blakeslee et al. 2003, Lidman et al.  2004, Mei
et al. 2006, Strazzullo et al. 2006).  Current results from the deep NIR
fields indicate a strong evolution in galaxy color segregation near redshift
1.7 (Cirasuolo et al. 2007). Deep X-ray surveys of the same fields are
therefore of further importance to provide the direct evidence of the role
of groups and clusters of galaxies in cosmic galaxy build up.

This paper concentrates on cataloging and analysis of the statistical
properties of the X-ray clusters primarily detected in XMM observations of
the Subaru XMM Deep Field. The basic X-ray data reduction and a construction
of the catalog of extended sources is discussed in \S\ref{data}. In
\S\ref{rs} we describe the cluster identification using a refined red
sequence method by including a galaxy preselection using multi-band
photometric redshift catalog. The stand of the spectroscopic follow-up is
presented in \S\ref{z}.  In \S\ref{xcat} we provide a final catalog of
identified clusters, including the results of the spectroscopic follow-up.
This is the first X-ray survey where special care is paid to select out the
systems where extended X-ray emission is caused by radio lobes. The details
and the results of this procedure are outlined in \S\ref{rg}. Statistical
properties of the clean X-ray cluster sample are discussed at the end of
\S\ref{xcat}.  \S\ref{resume} concludes the paper.

All through this paper, we adopt a ``concordance'' cosmological model, with
$H_o=72$ km s$^{-1}$ Mpc$^{-1}$, $\Omega_M=0.25$, $\Omega_\Lambda = 0.75$
(Komatsu et al. 2009), and --- unless specified --- quote all X-ray fluxes
in the [0.5--2] keV band and rest-frame luminosities in the [0.1--2.4] keV
band and provide confidence intervals at the 68\% level.

\begin{figure*}
\includegraphics[width=16.cm]{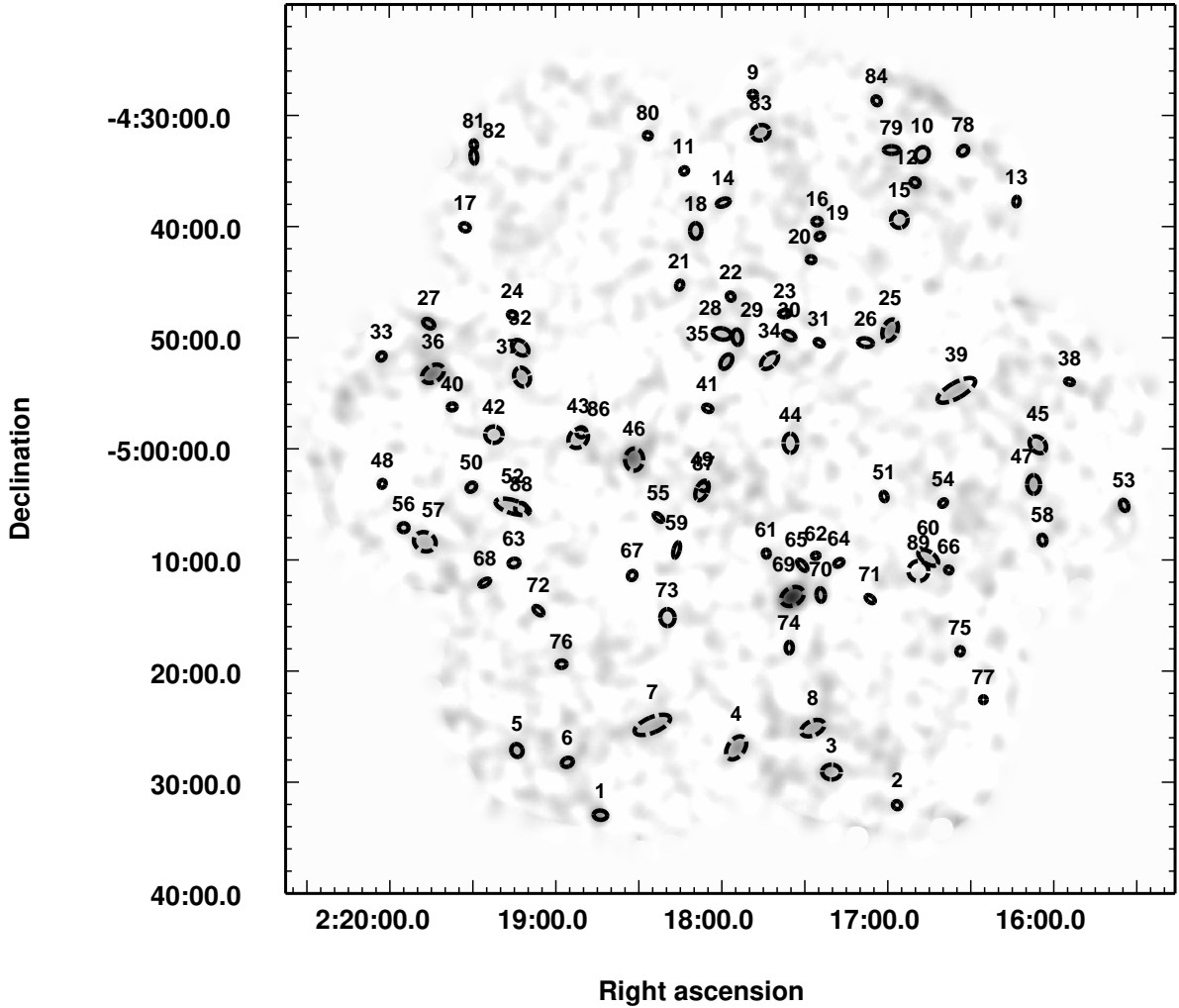}

\caption{An image of the signal-to-noise ratio in the 0.5--2 keV band
  after background subtraction and point source removal. The image has been
  smoothed with a Gaussian of $32^{\prime\prime}$ width. White color
  corresponds to the values smaller than 0, grey color starts at 1 sigma and
  black color corresponds to detection significance of 6 sigma per 200
  square arcsecond element. Ellipses indicate the wavelet sources, labeled
  according to the catalog. The coordinate grid is for the Equinox 2000. 
\label{f:s2n}}

\end{figure*}

\section{XMM Data reduction}\label{data}

The SXDF incorporates a deep, large-area X-ray mosaic with XMM-Newton,
consisting of seven overlapping pointings covering \mbox{1.3\sq\deg} region
of the high Galactic latitude sky with an exposure time of 100 ks in the
central field (in separate exposures) and 50 ks in the flanking fields (for
details see Geach et al. 2007). Four of the pointings were carried out in
2000 August, and the remaining three were made in August 2002 and January
2003.

For cluster detection, we used the XMM-Newton mosaic image in the 0.5--2 keV
band, consisting of 7 pointings, 400 ksec in total.  A description of the
XMM-Newton observatory is given by Jansen et al. (2001). In this paper we
use the data collected by the European Photon Imaging Cameras (EPIC): the
{\it pn}-CCD camera (Str\"uder et al. 2001) and the MOS-CCD cameras (Turner
et al. 2001).  All Epic-{\it pn} observations have been performed using the
Thin filter, while both Epic-MOS cameras used the Medium filter.

In addition to the standard data processing of the EPIC data, which was done
using XMMSAS version 6.5 (Watson et al. 2001; Kirsch et al. 2004; Saxton et
al. 2005), we perform a more conservative removal of time intervals affected
by solar flares, following the procedure described in Zhang et al. (2004).
In order to increase our capability of detecting extended, low surface
brightness features, we have applied the 'quadruple background subtraction'
(Finoguenov et al. 2007) and also check for high background that can be
present in a few MOS chips (Snowden et al. 2008), identifying none. The
resulting countrate-to-flux conversion in the 0.5--2 keV band excluding the
lines is $1.59\times10^{-12}$ for {\it pn} and $5.41\times10^{-12}$ for each
MOS detector, calculated for the source spectrum, corresponding to the APEC
(Smith et al. 2001) model for a collisional plasma of 2 keV temperature, 1/3
solar abundance and a redshift of 0.2. We note that in reconstructing the
properties of the identified groups and clusters of galaxies, we implement
the exact corrections, based on the source spectral shape (as defined by the
expected temperature of the emission) and the measured redshift of the
system.

After the background has been estimated for each observation and each
instrument separately, we produce the final mosaic of cleaned images and
correct it for the mosaic of the exposure maps in which we account for
differences in sensitivity between pn \& MOS detectors. 

We use the prescription of Finoguenov et al. (2009) for extended source
detection, which consists in removal the PSF model for each detected point
source from the data before applying the extended source search algorithm.
The signal-to-noise ratio image of the point-source cleaned image is shown
in Fig.~\ref{f:s2n}. As can be seen from the figure, the image exhibits a
fairly uniform signal-to-noise ratio.  Without the refined background
subtraction, the signal-to-noise image exhibited large-scale variations,
which could mimic an extended source. On the image, the ellipses show the
position and the angular extent of detected sources. The total number of
extended sources detected is 84. Identification of sources required to split
several sources, increasing the total to 92. The threshold for the wavelet
source detection has been set to 4 standard deviations. The calibrated map
of the wavelet noise (Vikhlinin et al. 1998) has been produced and used for
modelling of the survey sensitivity. The extent of the source, which we used
for identification and flux estimates, has been followed down to 1.6 times
the local wavelet noise value. The significance of the flux estimate can be
lower than 4 sigma. This is due to both a change in the significance of the
source between the peak of its significance and its extent as well as a
difference in the error field for detection and flux extraction. The later
difference is driven by difference in the fluctuation level between the
wavelet noise (important for detection) and unsmoothed source+noise
(important for the flux estimate). These differences decrease with the
increasing exposure of the survey (and e.g., are gone in our analysis of
CDFS (Finoguenov et al. in prep.).  Apart from the allowance for
systematical errors associated with AGN removal lately, the procedures of
calculating the sensitivity maps are the same as in Finoguenov et al.
(2007). Prior removal of point sources simplifies the extended flux estimate
and also allows us to use the X-ray center of extended emission as a prior
for identification, as detailed below. A small fraction of sources
($\sim15$\%) remains unidentified even in the area with best follow-up data.
Only 6\% can be accounted for by remaining deficiencies with the source
identification (see below). We believe this can be an effect associated with
joined detection of a number of sub-threshold point sources (e.g. Burenin et
al. 2007). A study of the origin of this source population is on-going using
CDFS and CDFN data, where one can profit from Chandra resolution (Finoguenov
et al. in prep.).

\begin{figure*}
\includegraphics[width=18.cm]{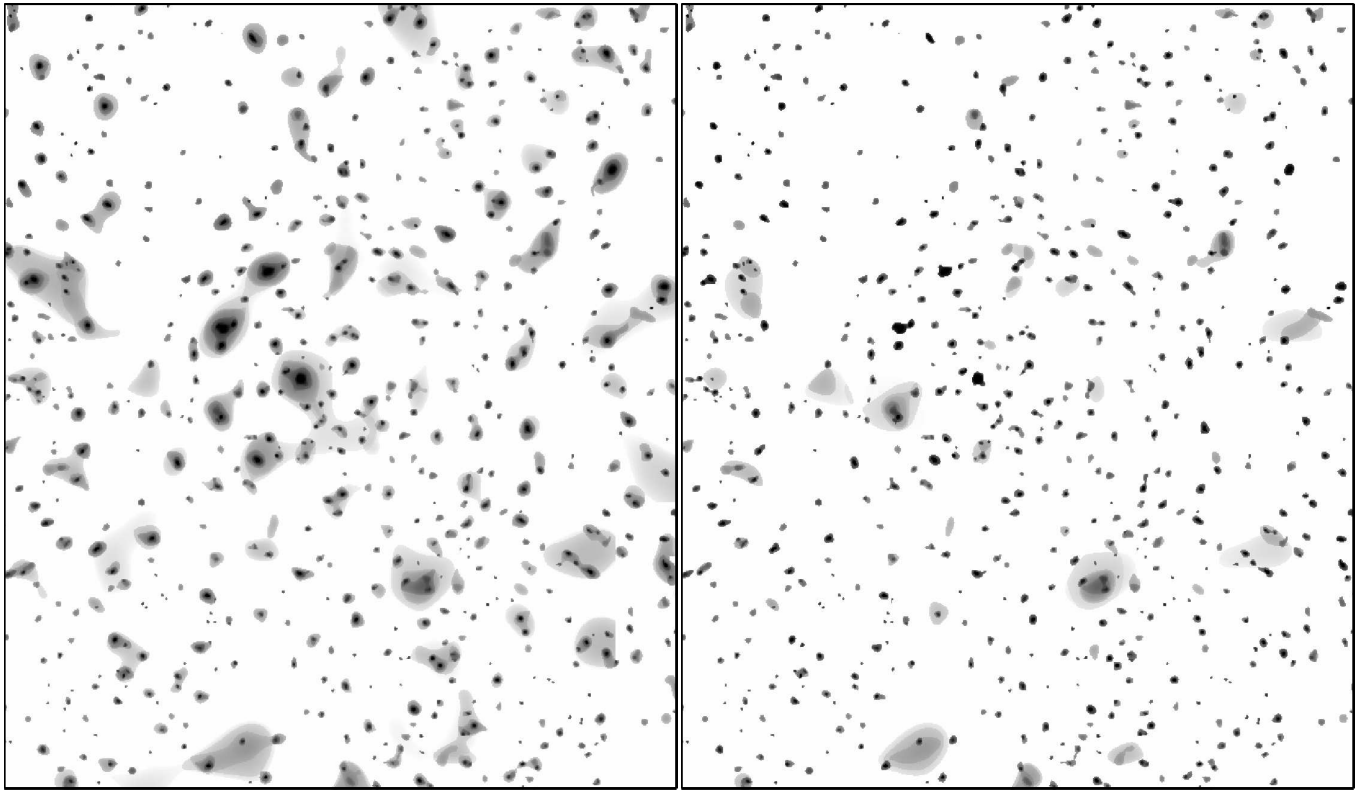} 
\caption{
Comparison of two wavelet reconstructions of the SXDF field. {\it Left panel}
displays a standard scale-wise decomposition, {\it right panel} includes a
three-level subtraction of the PSF wings associated with the point-like
sources, as described in the text. The scale limits for both images are the
same. Clearly the left panel is much less diffuse. Both images are $45^\prime\times55^\prime$. The pixel size is $4^{\prime\prime}$ on a side.
\label{f:x2ph}}

\end{figure*}

\section{Refined red sequence technique for cluster identification}\label{rs}

As a primary method for cluster identification we used the refined red
sequence technique, described in this section. This is a further refinement
of the photo-z concentration technique, used in Finoguenov et al. (2007).
Uncertainty, related to identification of clusters based on the photo-z data
alone has been addressed in van Breukelen et al. (2007). We deem our
technique as the most robust cluster identification when only broad band
photometric data are available. With the refinements, described here, this
technique is also sensitive to identification of galaxy groups. However,
strongly star-forming galaxy groups can not be detected through such
techniques. This point is thoroughly checked using zCOSMOS survey, yielding
only 1\% of such systems (Finoguenov et al. in prep.), which are located at
$z<0.3$. As we discuss below, our identification at $z<0.3$ is anyway
not complete, due to lack of U-band photometry.

First we consider the calibration of the model red sequence, then we detail
the application procedure and describe selection criteria. To model the
evolution of the red sequence, we adopt the passive evolution model of a
single stellar population (SSP), assuming no dust extinction, using the
Bruzual \& Charlot (2003) population synthesis code. In order to reproduce
the slope of the red sequence, the red sequence in the Coma cluster (Bower,
Lucey \& Ellis 1992) has been fitted by the SSP models formed at $z_f=5$
with various metallicities.  Model galaxies are 'calibrated' in this way
(Lidman et al.  2008). The model red sequence can then be evolved back in
time to arbitrary redshifts. Note that this modeling is based on the
assumption that the slope of the red sequence is entirely due to the
mass-metalicity relation, as suggested both by observations and in
theoretical work (e.g., Kodama \& Arimoto 1997, Stanford et al. 1998).

The fitting procedure is three-fold.  First, we extract galaxies in the area
centered on the extended X-ray emission. We then estimate significance of an
overdensity of red galaxies around the model red sequence at a given
redshift.  This procedure is performed at $0<z<2.5$.  All significant (with
details given in the following subsections) detections are stored for
further consideration.  As next we go through the step of approving the
identifications. The results of spectroscopic follow-up of similar sources
in CNOC2 field, discussed in Finoguenov et al. (2009), showed that the most
reliable identification has more than 3 galaxies inside the X-ray shape of
the source. In using the red sequence, to avoid chance projection a small
dispersion of galaxies with respect to the model red sequence shall also be
preferred.  We therefore favor these identifications, even if they are not
the most significant ones. When two or more identifications fit, we consider
splitting the X-ray source according to the galaxy counterpart and check
significance of these new sources, retaining only the significant ones and
assigning a lower flag (=2), if such a procedure is impossible, the X-ray
source is identified as confused (flag=4).  Robust identifications have a
clear concentration of red galaxies inside the detected X-ray emission. Some
identifications are less obvious and require more follow-up work. These are
marked correspondingly (flag=3).  A comprehensive list of source flags is
discussed in \S\ref{xcat}. Now we present the details of the method.

\subsection{Construction of the photometric catalog}

We use the Subaru $BVRiz$ photometric catalog from DR1 (Foucaud et al. 2007;
Furusawa et al. 2008).  The $z$-band selected catalog is used for this work.
We also use $JHK$ photometry from UKIDSS (UKIRT Infrared Deep Sky Survey)
Ultra Deep Survey (UDS) Third Data Release (Warren et al. 2009).  Objects
from UDS are cross-correlated with the Subaru catalog. The catalog is
further supplemented by Spitzer IRAC photometry from deep SpUDS program (PI
James Dunlop) and Spitzer Wide-area InfraRed Extragalactic survey (SWIRE,
Lonsdale et al. 2003).  We use SExtractor to detect objects on the IRAC
images and cross-correlate the IRAC objects with Subaru objects.  We use
$2^{\prime\prime}$ aperture for photometry and apply aperture correction.
Since data from different telescopes have different PSF sizes, we use total
magnitudes to derive colors. Stars are removed on the basis of their colors
and compactness.

We then feed the catalog to our photometric redshift code. A detailed
description of the code is given in Tanaka et al. (2008), but a brief
outline is given here. The code uses a library of templates based on Bruzual
\& Charlot (2003) models.  We assume the $\tau$ model to describe the star
formation histories of galaxies and allow $\tau$, dust extinction and
intergalactic extinction to vary. Each observed object is fitted with all
the templates and the best-fitting model is determined using the $\chi^2$
statistics. The quality of the photometric redshift estimate has been
compared to the spectroscopic redshifts, yielding 10\% outliers
($|z_{phot}-z_{spec}|>0.2$) and $0.03(1+z)$ uncertainty on redshift estimate
below $z_{spec}=4$. Dominant fraction of outliers is at $z<0.3$ and is due
to lack of U-band data.

\subsection{Selection of Galaxies for cluster identification}

As mentioned above, we go over the redshift ($z$) at which we apply the
cluster red sequence method. We select galaxies at $|z-z_{phot}|<0.2$, where
$z_{phot}$ is a photometric redshift. As next, we only consider the galaxies
located within 0.5 Mpc (physical) from the center of X-ray emission at a
given redshift (see description of Eq.1 below for more details of the
weighting scheme). The aperture size is fixed on a physical scale and thus
its apparent size on the sky varies with redshifts, at which we look for an
over-density of red galaxies.  This radius is wide enough to include most of
the galaxies in a candidate cluster, while it is small enough to detect weak
signals from high redshift clusters.

Using a fixed aperture to select galaxies for the red sequence test is
sufficient for our purposes as the probed mass range of systems in the
survey is narrow and 0.5 Mpc typically encompasses $r_{500}$ of the system.
Alternative choice of galaxy selection can be either X-ray extent or an
estimate of $r_{500}$ based on the redshift guess and X-ray properties of
the system. The X-ray extent is determined by statistical significance of
the detection and would introduce uneven demand on matching between galaxies
and the X-ray source. Furthermore, for nearby objects the extent of the
emission is predicted to go into the scales where confusion becomes
important ($2^{\prime}$ for the depth of our survey), so the observed extent
will be truncated. Using the fixed aperture, we can make a fair comparison
of significance level of detection of overdensities at various redshifts.

\subsection{Application of the red sequence method}

To probe if there is any overdensity of red galaxies at a given redshift
$z$, we count galaxies around the model red sequence.  We use a Gaussian
weight when counting galaxies in a form of

$$
\sum_i \exp\left [ -\left
    (\frac{color_{i,obs}-color_{model}(z)}{\sigma_{i,obs}}\right )^2\right
]$$
$$
 \times\exp \left [-\left (
     \frac{mag_{i,obs}-mag^*_{model}(z)}{\sigma_{mag}}\right) ^2\right
 ]\times 
$$
$$
\exp \left( -(\frac{r_i}{\sigma_r})^2\right) ,
\eqno (1)
$$


\noindent
where $color_{i,obs}$ and $mag_{i,obs}$ are the color and the magnitude of
$i$-th observed galaxy, $\sigma_{i,obs}$ is the observed color error in
$color_{i,obs}$, $color_{model}(z)$ is the model red sequence color at the
magnitude of the observed galaxy, $mag^*_{model}(z)$ is the characteristic
magnitude based on the model, which is tuned to roughly reproduce the
observed characteristic magnitudes, $\sigma_{mag}$ is the smoothing
parameter detailed below, $r_i$ is the distance from the X-ray center and
$\sigma_r$ is also a smoothing parameter as shown below.  To account for 
systematic zero point errors in observations and for systematic
magnitude/color errors in models, we take a minimum error in
$\sigma_{i,obs}$ of 0.1 mag.  For example, if an object has
$\sigma_{i,obs}<0.1$, we take $\sigma_{i,obs}=0.1$ for this object.

Since different colors are sensitive to red galaxies at different redshifts,
we adopt the following combination of colors and magnitudes.

\noindent
$0.0<z<0.5$ : $B-i$ color and $i$ magnitude\\
$0.5<z<1.0$ : $R-z$ color and $z$ magnitude\\
$1.0<z<1.5$ : $i-K$ color and $K$ magnitude\\
$1.5<z<2.5$ : $z-3.6\mu m$ color and $3.6\mu m$ magnitude\\

\noindent
In the Subaru/XMM Deep field, we have no $U$-band photometry, which is
crucial for low-$z$ red galaxies.  Thus, our method is not very sensitive to
clusters at $z<0.3$ with estimated incompleteness in the cluster catalog of
$\sim 6$\%.

The luminosity function of red galaxies varies with both richness of a
cluster and redshift (Tanaka et al. 2005, 2007). To minimize the richness
and redshift dependency of the red sequence technique, we weight galaxies
according to their luminosity. This is implemented in the second term of Eq.
1, by adding high weight to the detection of bright red galaxies, adjusted
according to passive evolution model, and the smoothing parameter
$\sigma_{mag}$, which we set to a value of 2.

The third term in Eq. 1 takes into account the concentrations of galaxies. A
galaxy at the center of the cluster has heavier weight than that in the
outskirts. The relative weight as a function of distance from the center is
controlled by the smoothing parameter $\sigma_r$.  We take $\sigma_r=1.0$
Mpc.  This means that a galaxy at 0.5 Mpc from the center (i.e., galaxy at
the edge of the extraction aperture) has a $e^{-0.25}=0.78$ weight relative
to a galaxy at the center. Altogether, we take into account the color
evolution and the magnitude evolution of the red sequence (1st and 2nd term,
respectively), the concentration of galaxies (3rd term) and a density of red
galaxies around the red sequence at any given redshift.

To quantify the significance of the overdensity of red galaxies at the
position of a cluster candidate, we put an aperture of the same size at a
random position in the Subaru-XMM Deep Field and perform the same procedure
500,000 times.  This gives an average number of red galaxies and its
dispersion in the field at a given redshift.  Then, the significance is
evaluated as a relative overdensity of the cluster candidate to that of the
field. A formal error of the red sequence redshift is not straight forward
to estimate since we use many parameters to derive the significance and the
absolute significance changes with these parameters.  To remedy this
complexity, we simply take the full-width at half maximum of the
significance peak as the error. In Tab.\ref{t:ol} we list the significance
of the red sequence and the uncertainty of the redshift of the cluster.

\section{Spectroscopic follow-up}\label{z}

Since 2004 the $z<1.3$ galaxy cluster candidates have been intensively
followed up as part of a SXDF VLA program on VLT and targeted Gemini
proposals (Simpson et al. 2006 and in prep.).  Geach et al.  (2007) reported
multi-object spectroscopy on four candidate X-ray galaxy groups around
moderate-luminosity radio sources. van Breukelen et al. (2007; 2009) report
some of the results of Gemini program.  Other spectroscopic observations of
the field are reported in Yamada et al.  (2005), Smail et al. (2008) and
Akiyama et al. (2dF/AAOmega, in prep.).  A total of 4k spectra have been
obtained thus far.

Using the identification of the cluster with a red sequence method, we
looked for spectroscopic redshifts of any of the red sequence galaxies.
Where there has been a consistent spectroscopic redshift found, we
considered it as a refinement. Next, we searched for more spectroscopic
redshifts in all galaxies matching the selected redshift to $0.005\times
(1+z)$, which is twice a typical velocity dispersion. In addition we have
also looked for galaxy clustering at different redshifts, when data allowed,
but found no outliers. In the cluster catalog we report both the
spectroscopic redshift when available and a number of galaxies used to
derive it, which can be used to assess the quality of the spectroscopic
follow-up.

In Tab.\ref{t:zgal} we list the coordinates and redshifts of the 144
galaxies assigned to the X-ray clusters. In case the cluster has less than 3
spectroscopic members, this assignment is tentative.

\section{A catalog of identified X-ray clusters}\label{xcat}

In this section we describe our catalog of 57 X-ray galaxy clusters detected
in the SXDF/UDS field. In the catalog (Tab.\ref{t:ol}) we provide the
cluster identification number (column 1), R.A. and Decl. of the peak of the
galaxy concentration identified with the extended X-ray source in Equinox
J2000.0 (2--3), photometric redshift (4). In case there are spectroscopic
redshift determination for the cluster member galaxies, the median
spectroscopic redshift is listed instead. The cluster flux in the 0.5--2 keV
band is listed in column 5 with the corresponding 1 sigma errors. The flux
has units of $10^{-14}$ ergs cm$^{-2}$ s$^{-1}$. The rest-frame luminosity
in the 0.1--2.4 keV band in units of $10^{42}$ ergs s$^{-1}$ is given in
(8).  Column 9 lists the estimated total mass, $M_{200}$, computed following
Rykoff et al.  (2008) and assuming a standard evolution of scaling
relations: $M_{200} E_z = f(L_X E_z^{-1})$. The corresponding $R_{200}$ in
arcminutes is given in column 10.  Column 11 lists X-ray flag and the number
of member galaxies inside $R_{200}$, $N(z)$ is given in column 12. The
errors provided on the derived properties are only statistical and do not
include the intrinsic scatter in the $L_X-M$ relation, which makes
individual mass estimates uncertain by 0.2dex (Vikhlinin et al. 2009).  To
provide the insights on the reliability of both the source detection and the
identification, in col (13) we provide the significance of the X-ray flux
estimate and in col (14) the significance of the red sequence.  Col. (15)
shows the red sequence redshift and its uncertainty. Col. (16) provides the
median photometric redshift of galaxies on red sequence.  Finally, col. (17)
provides a reference to the extended source catalog in Ueda et al. (2008).

Both flux estimates and the calculation of the properties of clusters are
similar to the procedure outlined in Finoguenov et al. (2007).  An X-ray
quality flag ({\sc xflag}) have been derived for the entire catalog based on
visual inspection. {\sc xflag=1} is assigned to objects with high (in
general $>6$) significance of the X-ray flux estimate, and having a single
optical counterpart. The next category of clusters ({\sc xflag=2}), are low
significance detections, for which X-ray centering has a large uncertainty
(up to $30^{\prime\prime}$), hence a larger weight is given to the location
of the optical counterpart. In addition, in cases when a single X-ray source
has been split into several sources, matching the optical counterpart, the
assigned flag is set equal 2 or larger. The clusters for which the
photometric redshift of the optical or NIR counterpart is uncertain are
flagged as {\sc xflag=3}, which is mostly a concern at $z>1.2$. {\sc
  xflag=4} indicates a presence of multiple optical counterparts, whose
contribution to the observed X-ray emission is not possible to separate or
rule out. Finally, the systems with potentially wrong assignment of an
optical counterpart are marked as {\sc xflag=5}.

In Ueda et al. (2008), the results of the analysis of the same XMM data has
been presented, identifying a total of 32 extended sources. The extended
sources were identified by examining source extent over the point spread
function assuming a Gaussian as the intrinsic image profile.  For extended
sources considered in this paper, we search for sources from Ueda et al.
(2008) catalog around the cluster central position within a radius of 32
arcsec, a typical size of significant extended emission. The positional
errors (1$\sigma$) of the Ueda et al. (2008) sources are also taken into
account in the matching. We recover 20 out of the 32 extended sources in
Ueda et al.  (2008) and identify 15 of them as clusters.  In Tab.\ref{t:ol}
we provide a match between our cluster catalog and the extended sources in
Ueda et al. (2008). The remaining differences can be understood as different
sensitivities of the algorithm to extent of X-ray emission, which in our
case is taken into account in the modelling of the survey.
%

\begin{figure}
\includegraphics[width=7.5cm]{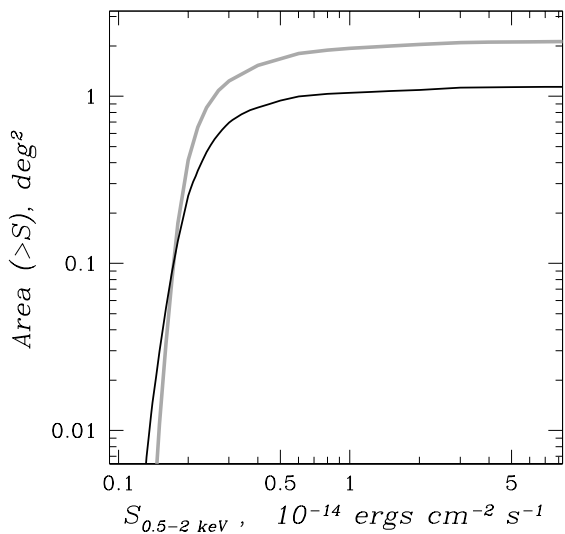}

\caption{Survey area of SXDF (black curve) as a function of the total source
flux in the 0.5--2 keV band. COSMOS flux-area curve corresponding to the
results in Finoguenov et al. (2007) is shown as grey line.   \label{f:area}}
\end{figure}

\begin{figure}
\includegraphics[width=7.5cm]{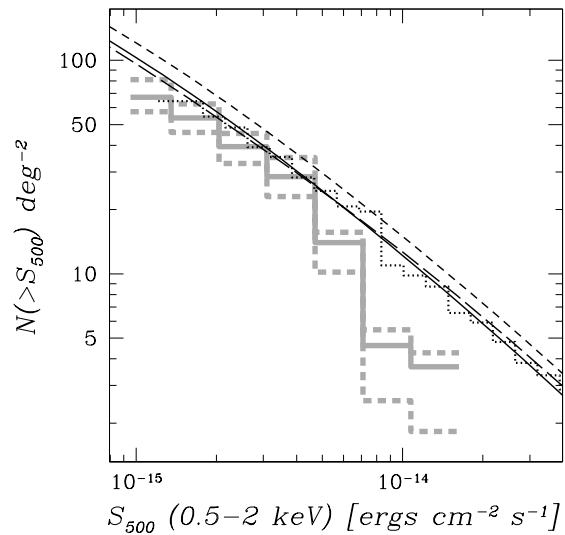}

\caption{Cumulative cluster number counts ($\log(N>S)-\log(S)$) for the SXDF
  field. The solid grey histogram shows the data and dashed grey histograms
  denote the 68\% confidence interval. The dotted histogram shows a
  $\log(N>S)-\log(S)$ of the COSMOS survey. The long dashed curve shows the
  prediction for no evolution in the luminosity function in Rosati et al.
  (2002), which provides a good fit to the date above $10^{-14}$ ergs
  s$^{-1}$ cm$^{-2}$ flux. The dashed line shows the WMAP5 predictions for
  $\log(N>S)-\log(S)$ under our assumptions for scaling relations and their
  evolution excluding. The solid line has been produced to match
  observational predictions by adopting a 5\% reduction in $\sigma_8$.
  \label{f:logn}}
\end{figure} 

\subsection{Identification of X-ray jets and halo occupation statistics of
  radio-galaxies}\label{rg} Extensive Inverse Compton X-ray
emission from large radio galaxies has been detected above redshift 1, such
as 3C~356 (z=1.12; Simpson \& Rawlings 2002), 3C~294 (z=1.786; Fabian et al.
2003), 6C~0905+39 (z=1.833 Erlund et al. 2008) and 4C~23.56 (z=2.48; Johnson
et al. 2007). The flux of the emission depends on the energy density of the
target photons, which in case of the CMB rises as $(1+z)^4$ so canceling out
the dimming expected from increased distance (Felten \& Rees 1969; Schwartz
2002). Those studies predict a large number of extended sources detected in
deep X-ray surveys, whose origin of the emission does not stem from the hot
gas associated with potential wells of those systems, but is instead caused
by the inverse Compton scattering of CMB photons on the relativistic
electrons of a Mpc jet. However, there has not been a single survey, which
can quantify the effect.

\begin{figure*} \mbox{\includegraphics[width=6.cm]{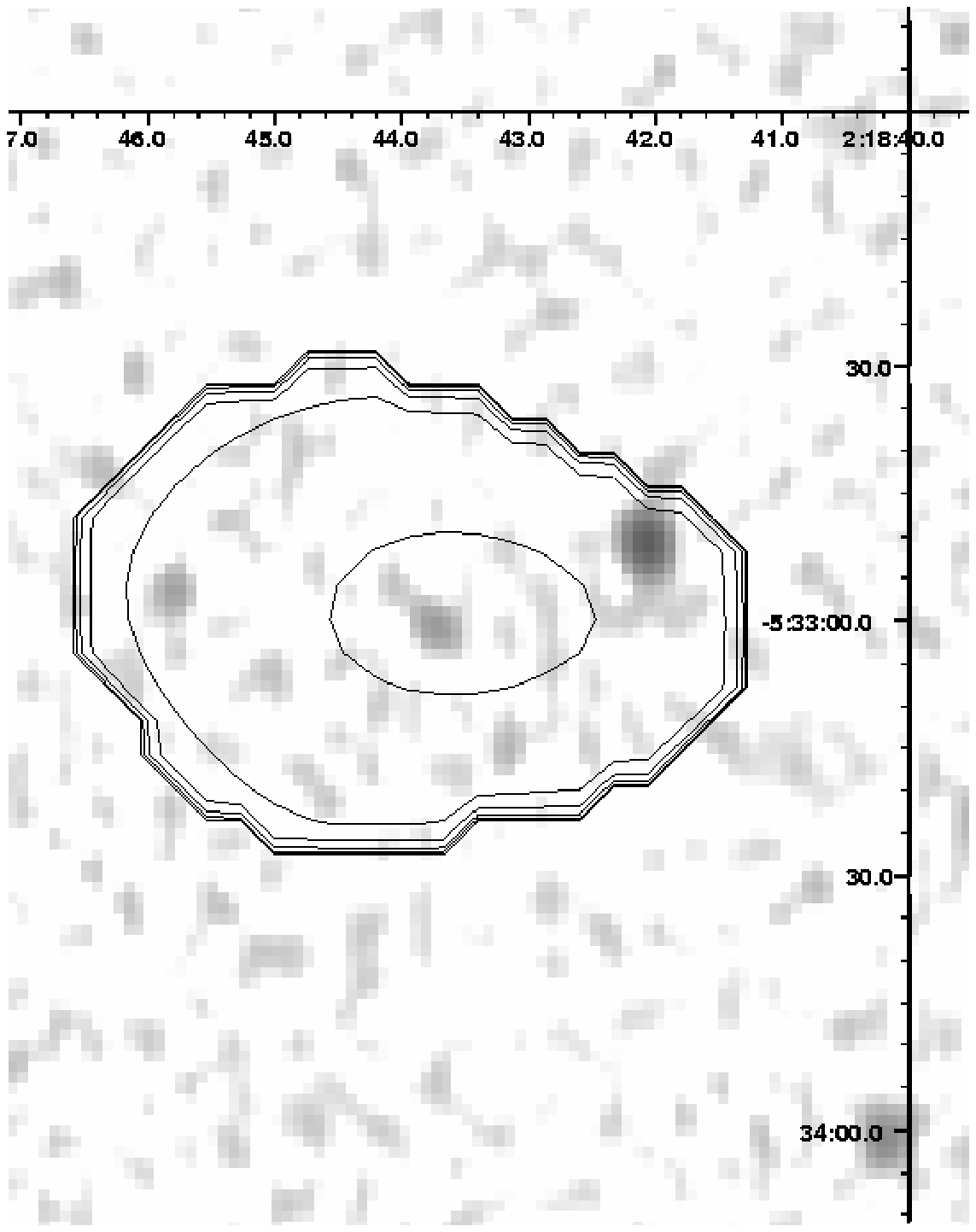}
    \hfill \includegraphics[width=6.cm]{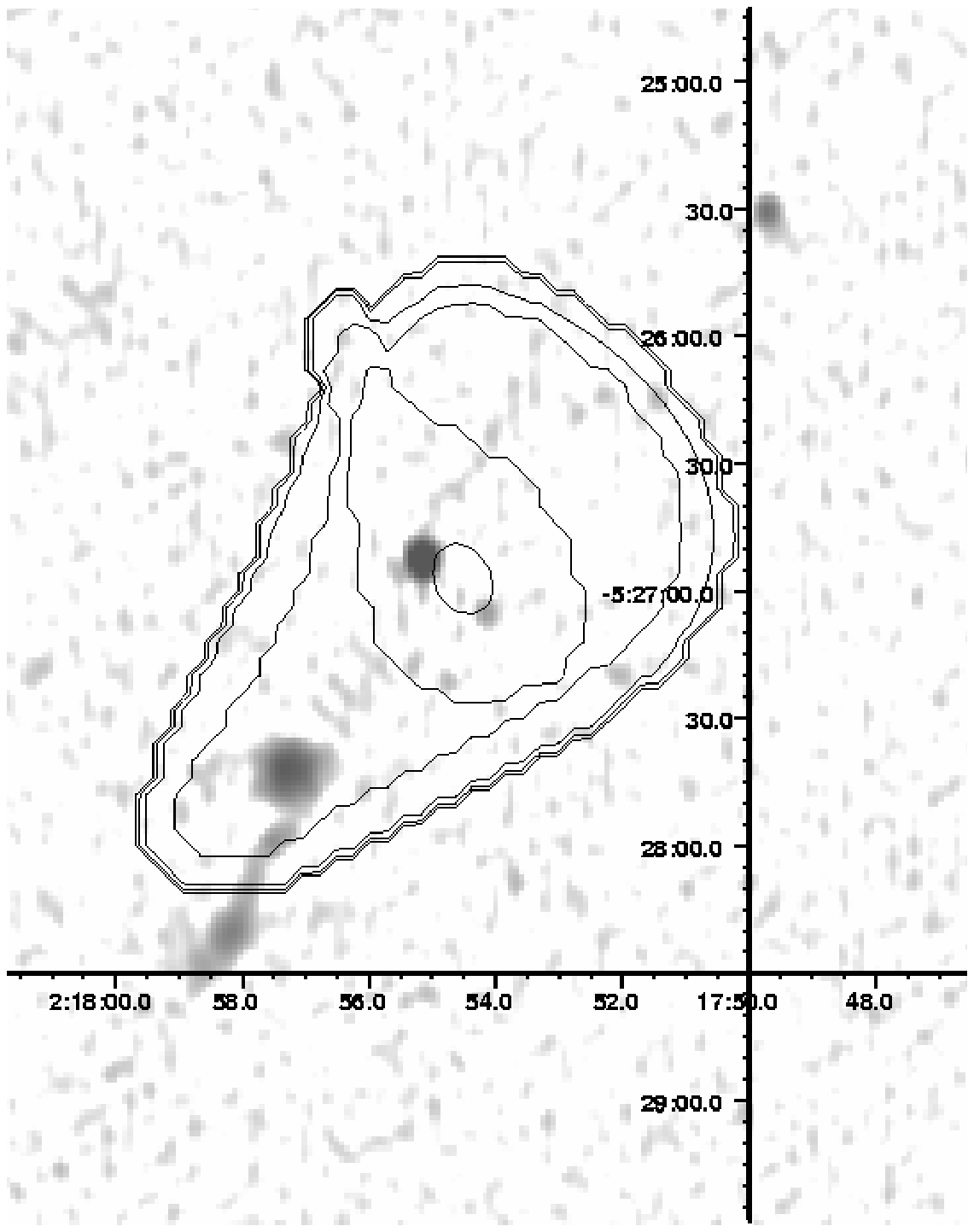} \hfill
    \includegraphics[width=6.cm]{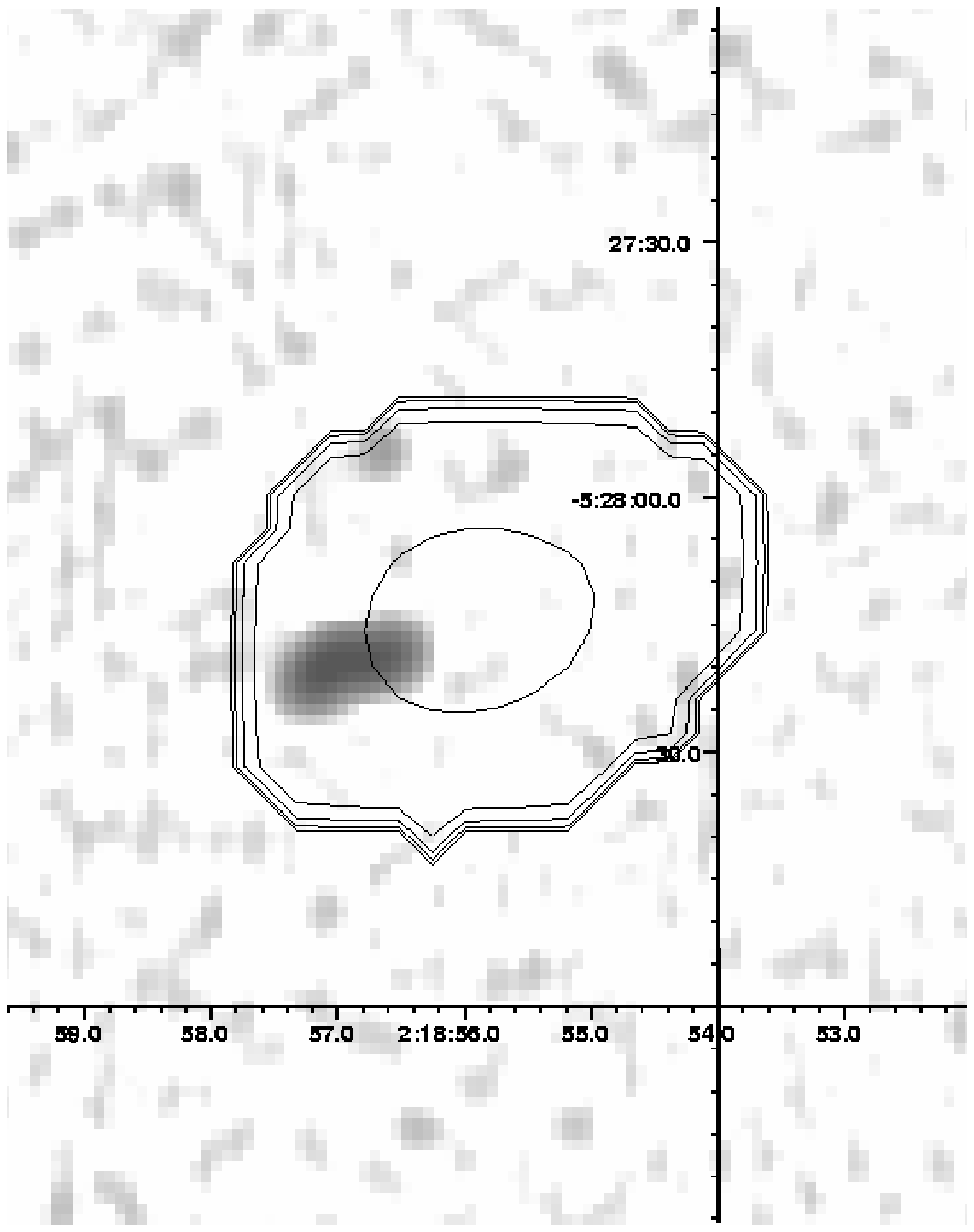}} \\
  \mbox{\includegraphics[width=6.cm]{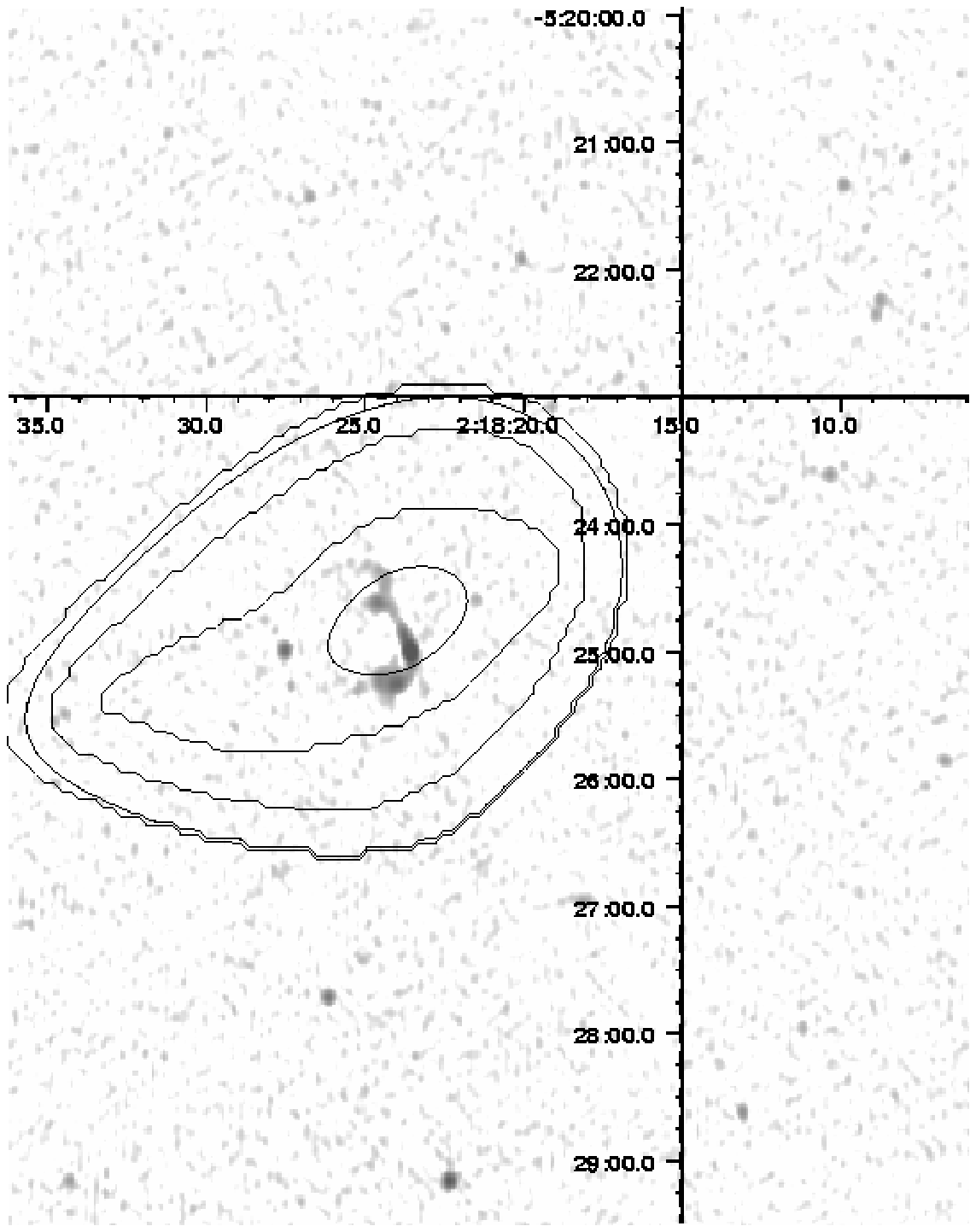} \hfill
    \includegraphics[width=6.cm]{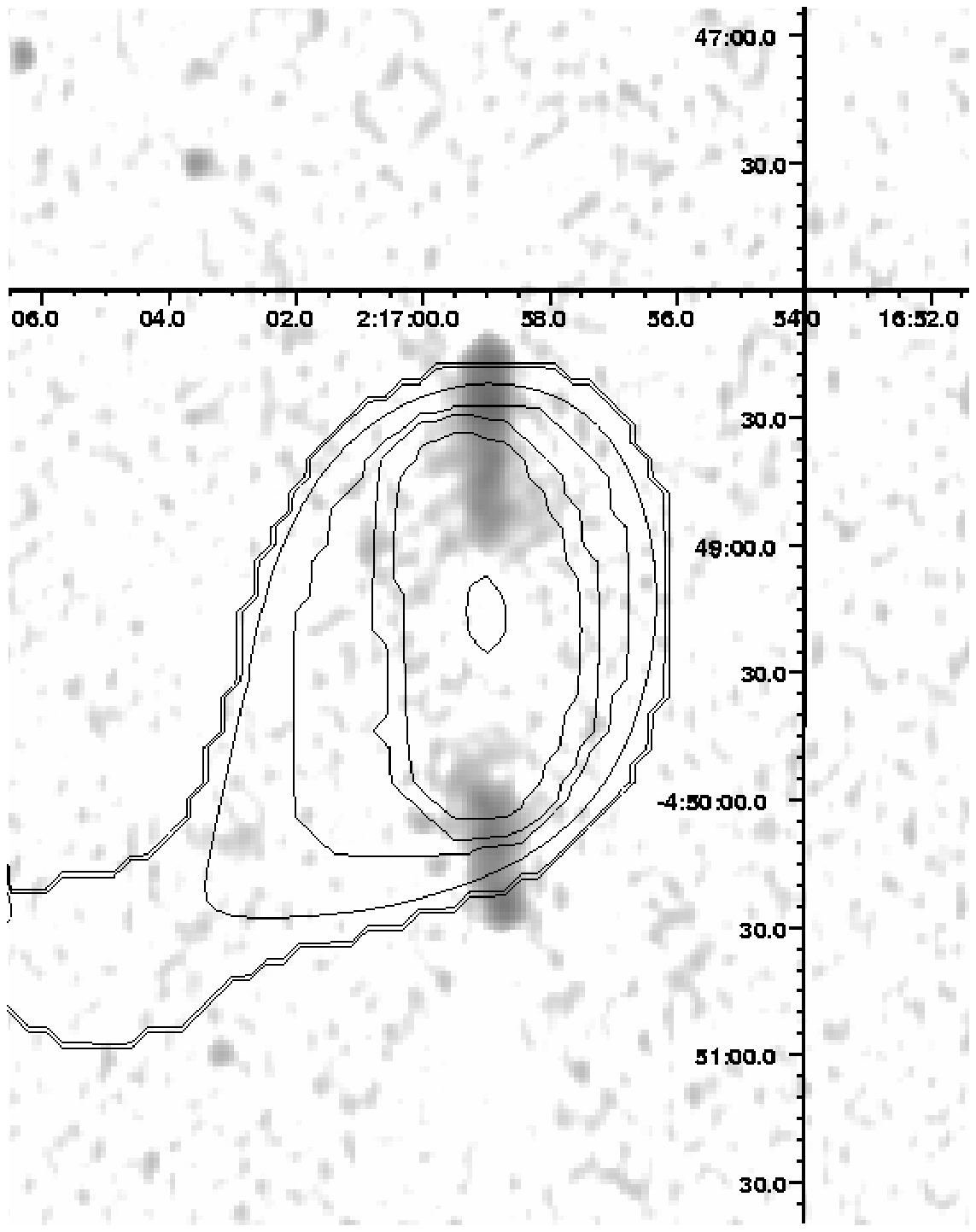} \hfill
    \includegraphics[width=6.cm]{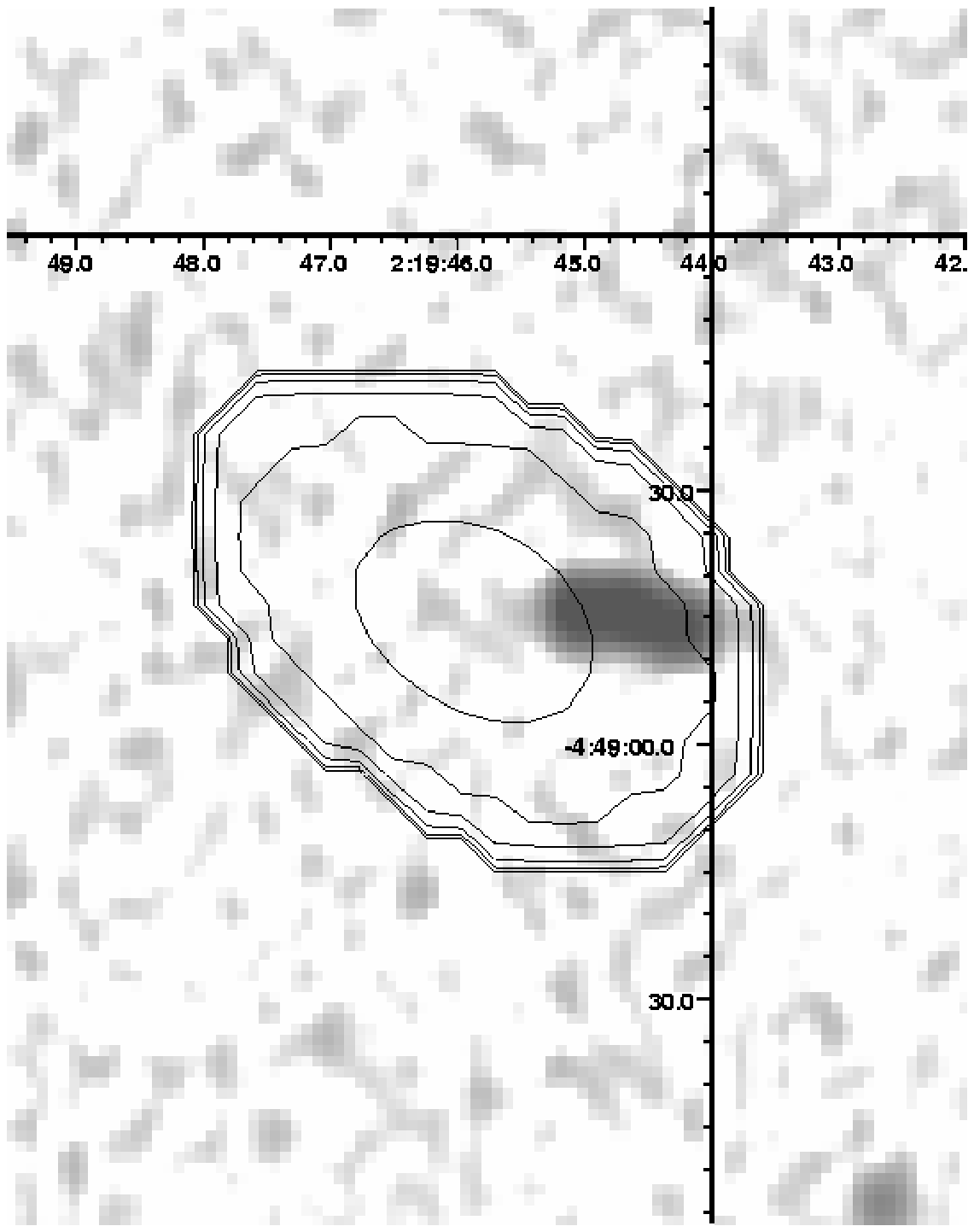}} \\
  \mbox{\includegraphics[width=6.cm]{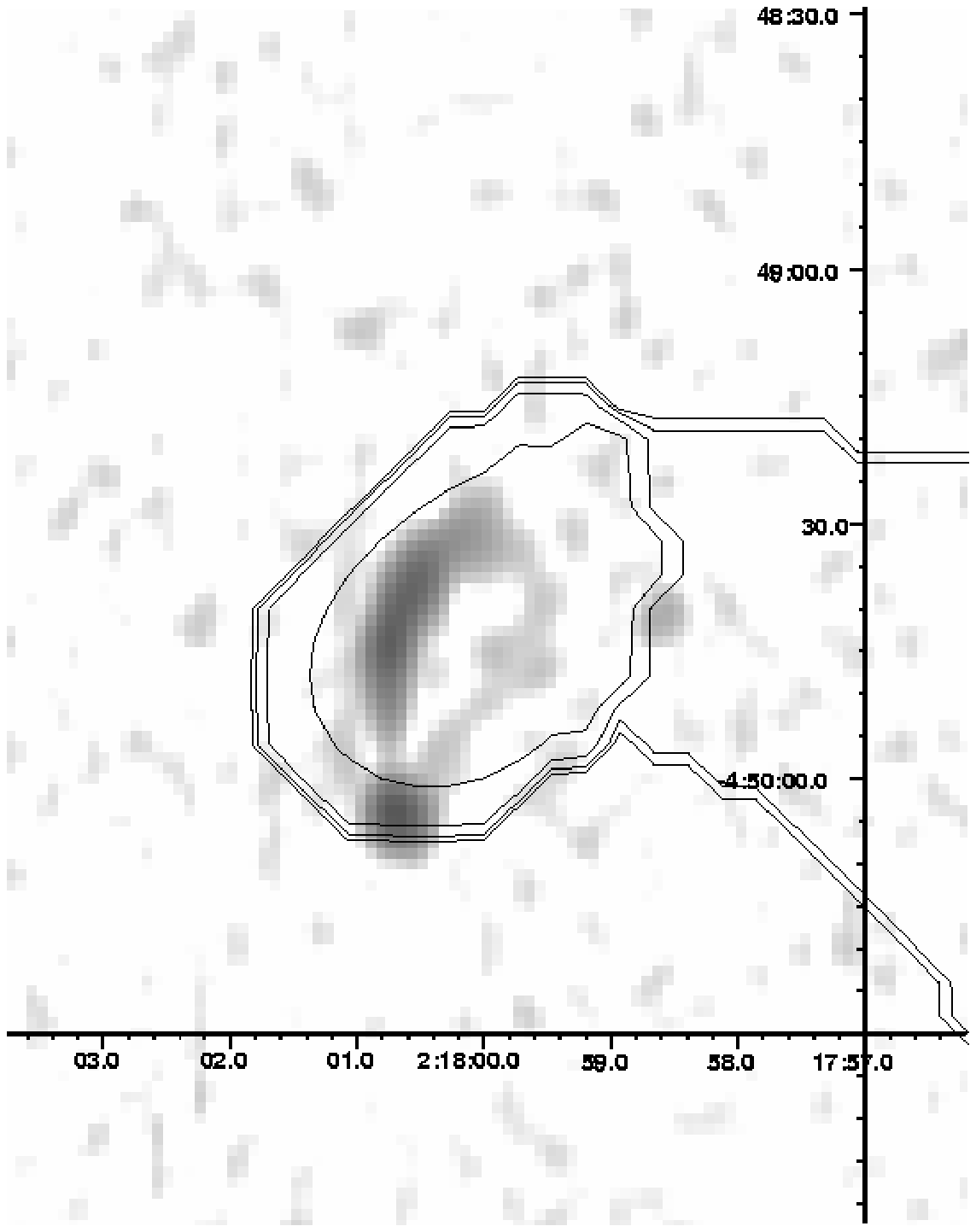} \hfill
    \includegraphics[width=6.cm]{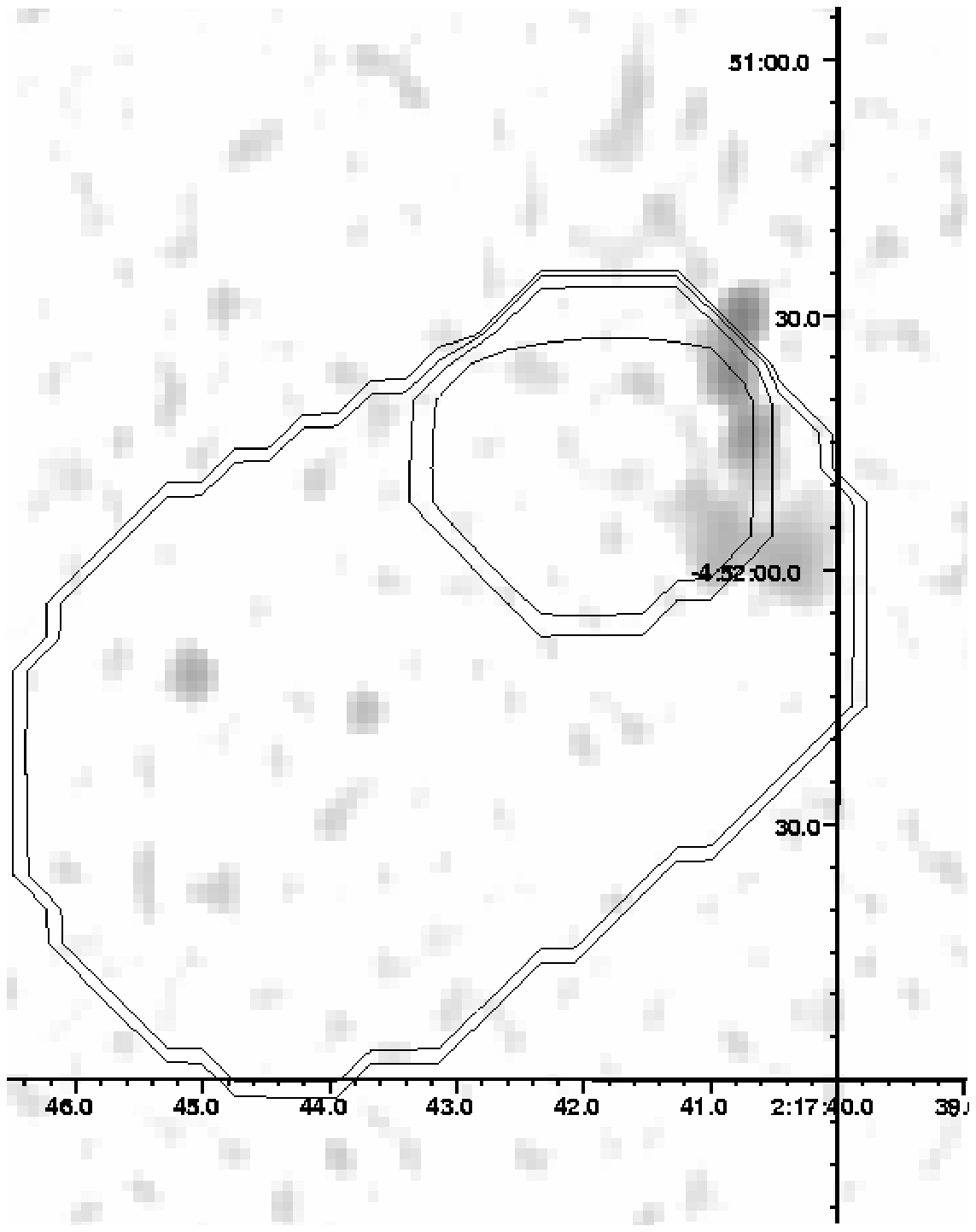} \hfill
    \includegraphics[width=6.cm]{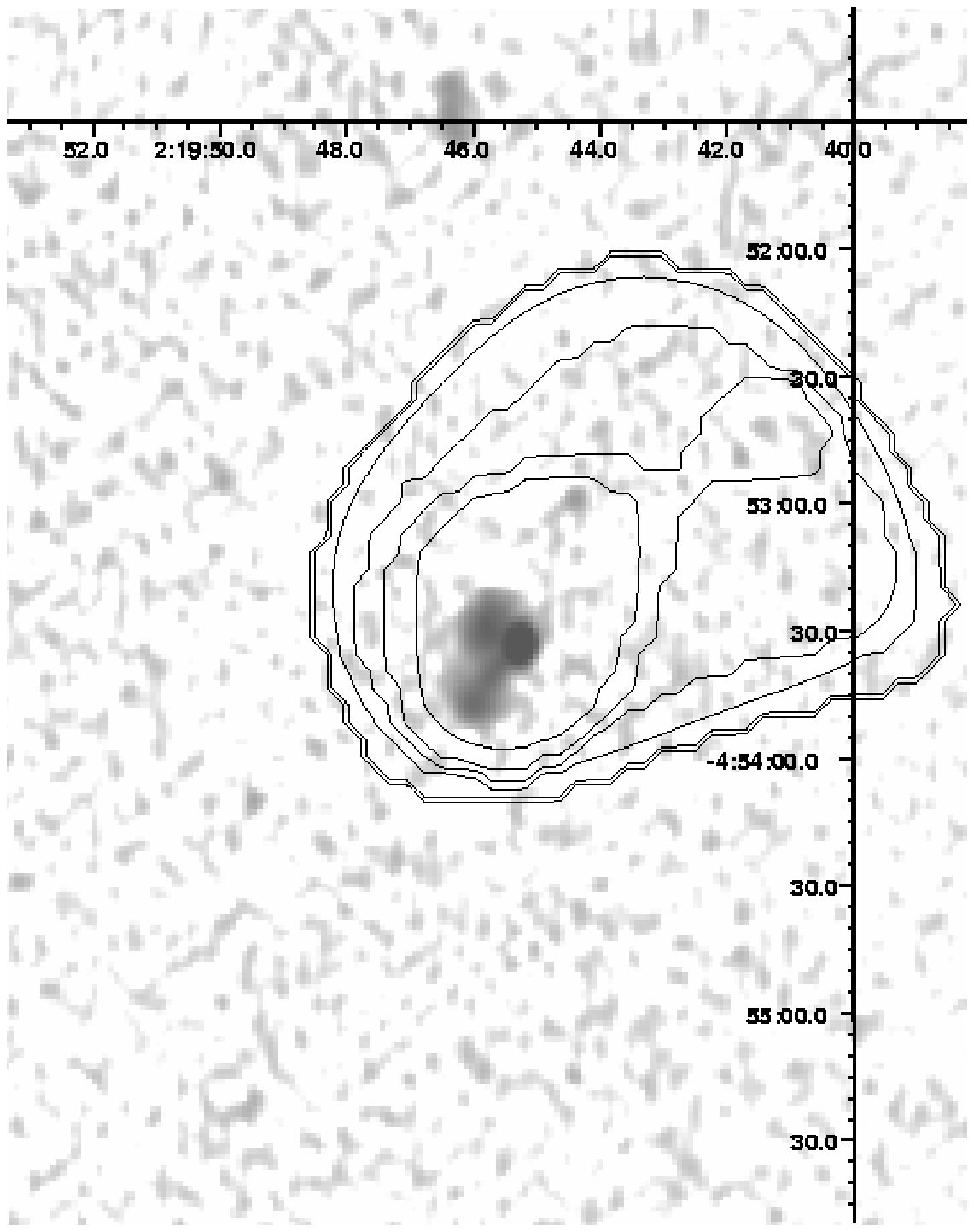} } \caption{Comparison
    of complex radio sources with extended X-ray emission in SXDF. 
      Images at 1.4 GHz frequency are overlaid with contours showing the
      wavelet reconstruction of extended
      X-ray emission 
    on spatial scales of 32$^{\prime\prime}$, 64$^{\prime\prime}$, and
    128$^{\prime\prime}$. From upper left to lower right the cluster IDs are
    1,4,6,7,25,27,28,34,36.  The coordinate grid is for the Equinox
    2000.  The astrometric differences shall be of the order of
    $2^\prime\prime$ and do not matter for this comparison.
  \label{f:radio}}
\end{figure*} \begin{figure*} \addtocounter{figure}{-1}
  \mbox{\includegraphics[width=6.cm]{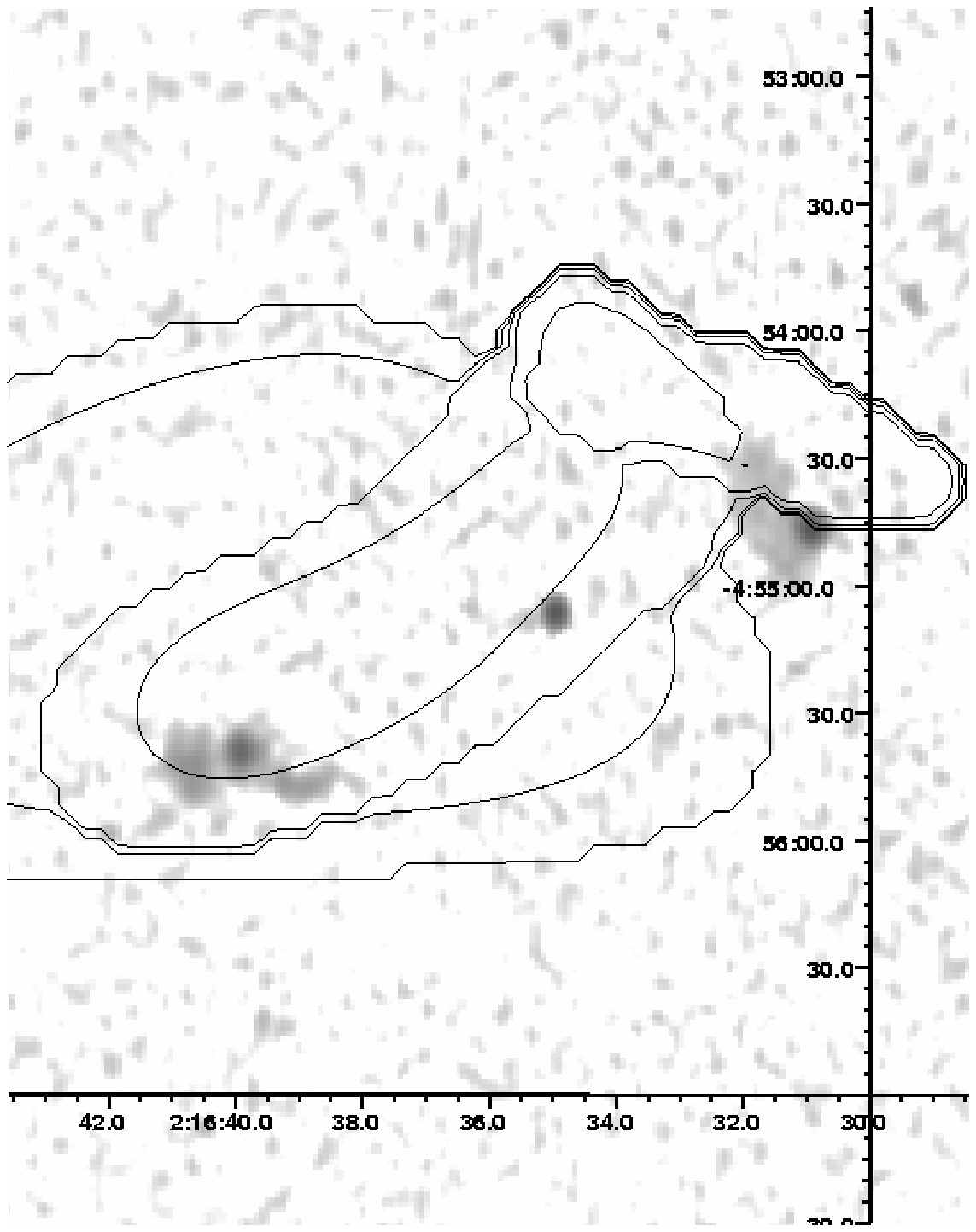} \hfill
    \includegraphics[width=6.cm]{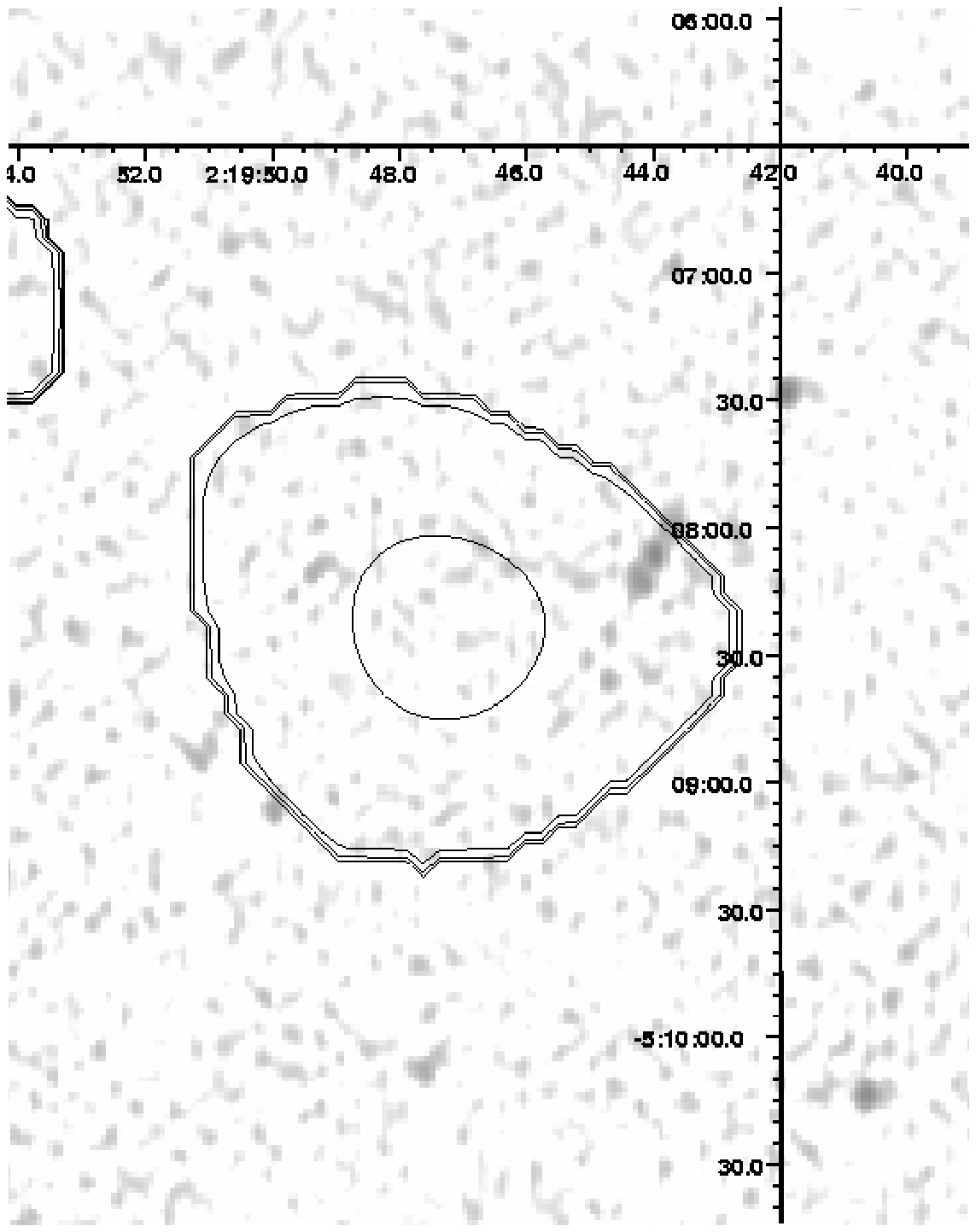} \hfill
    \includegraphics[width=6.cm]{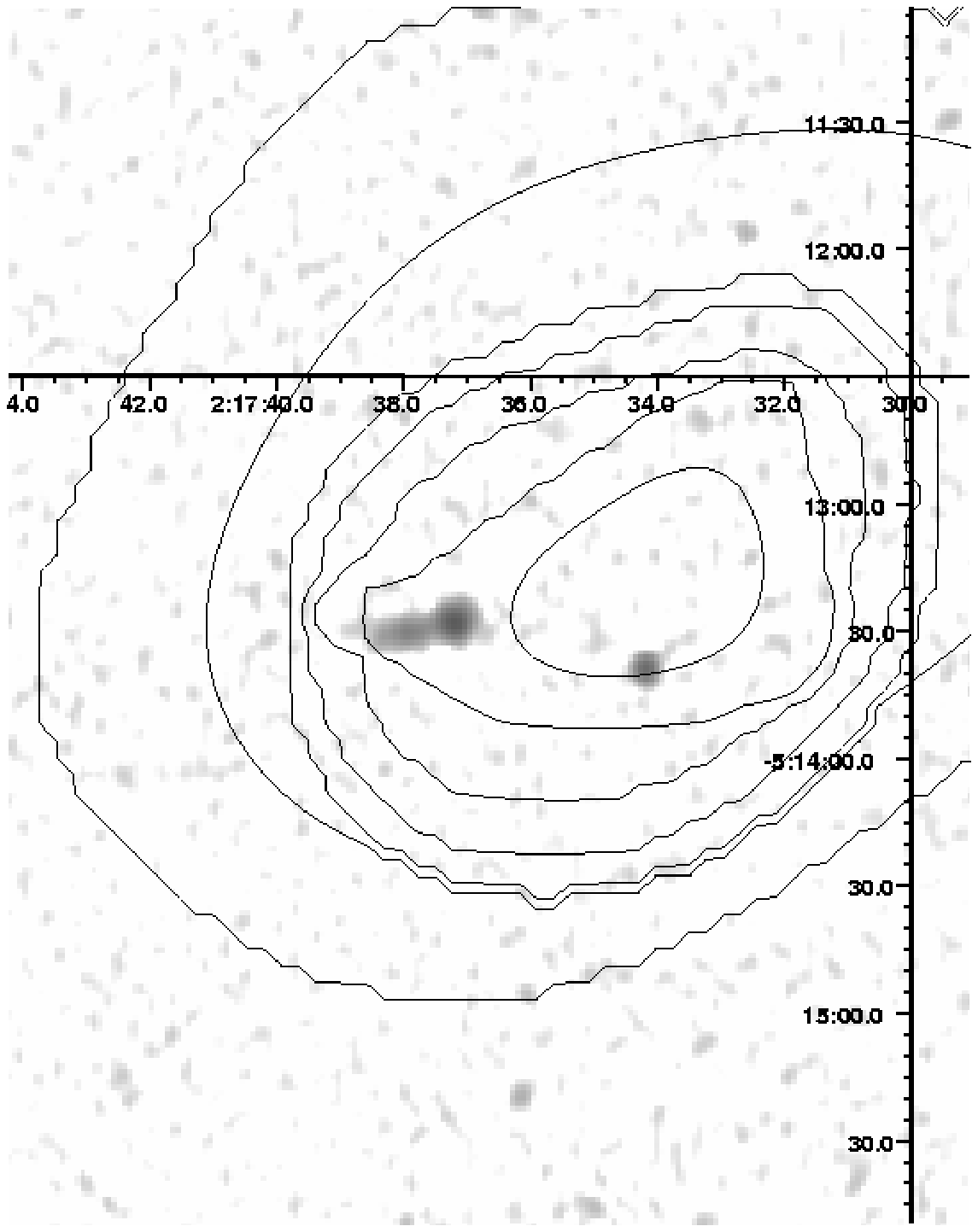} } \\
  \mbox{\includegraphics[width=6.cm]{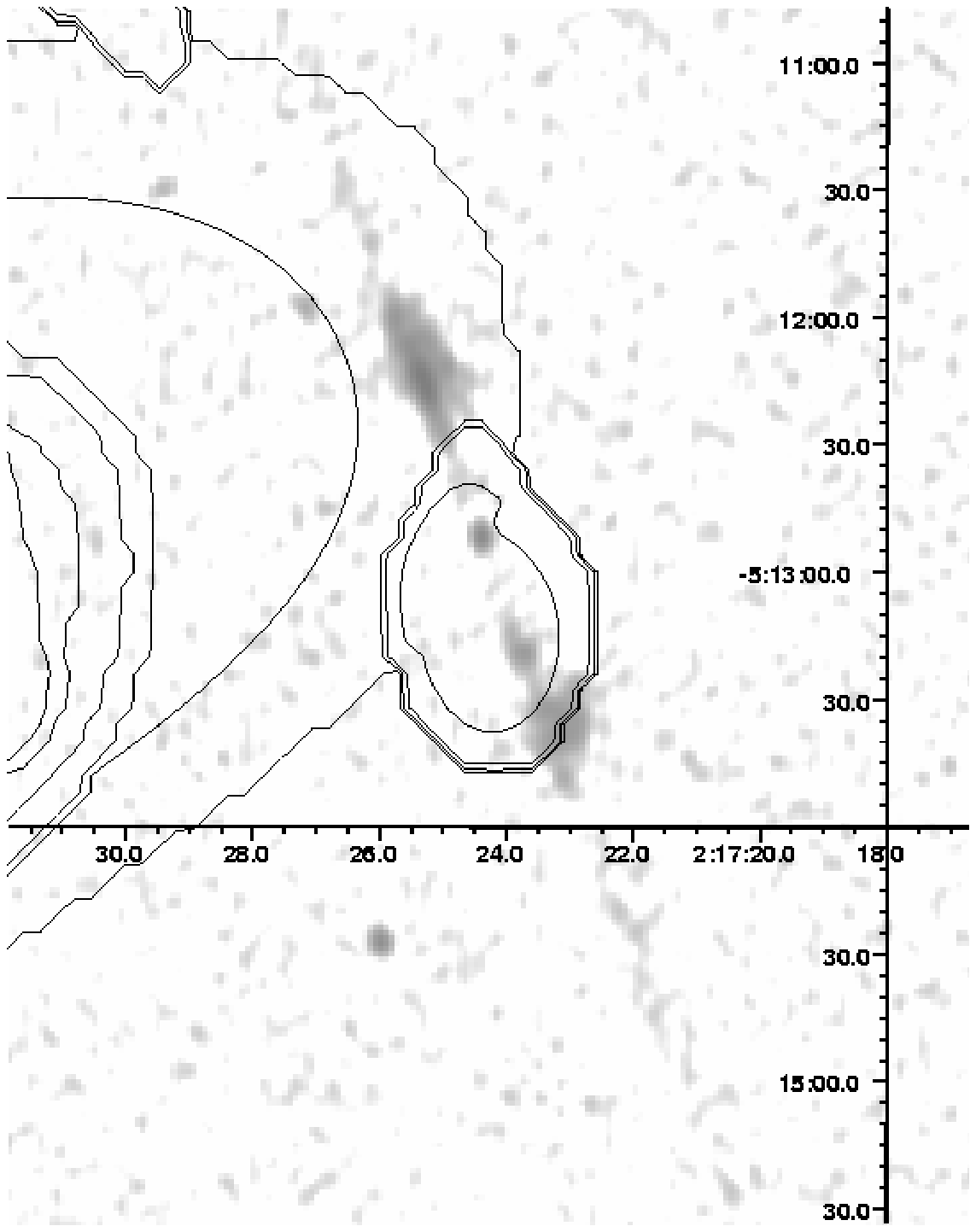} \hfill
    \includegraphics[width=6.cm]{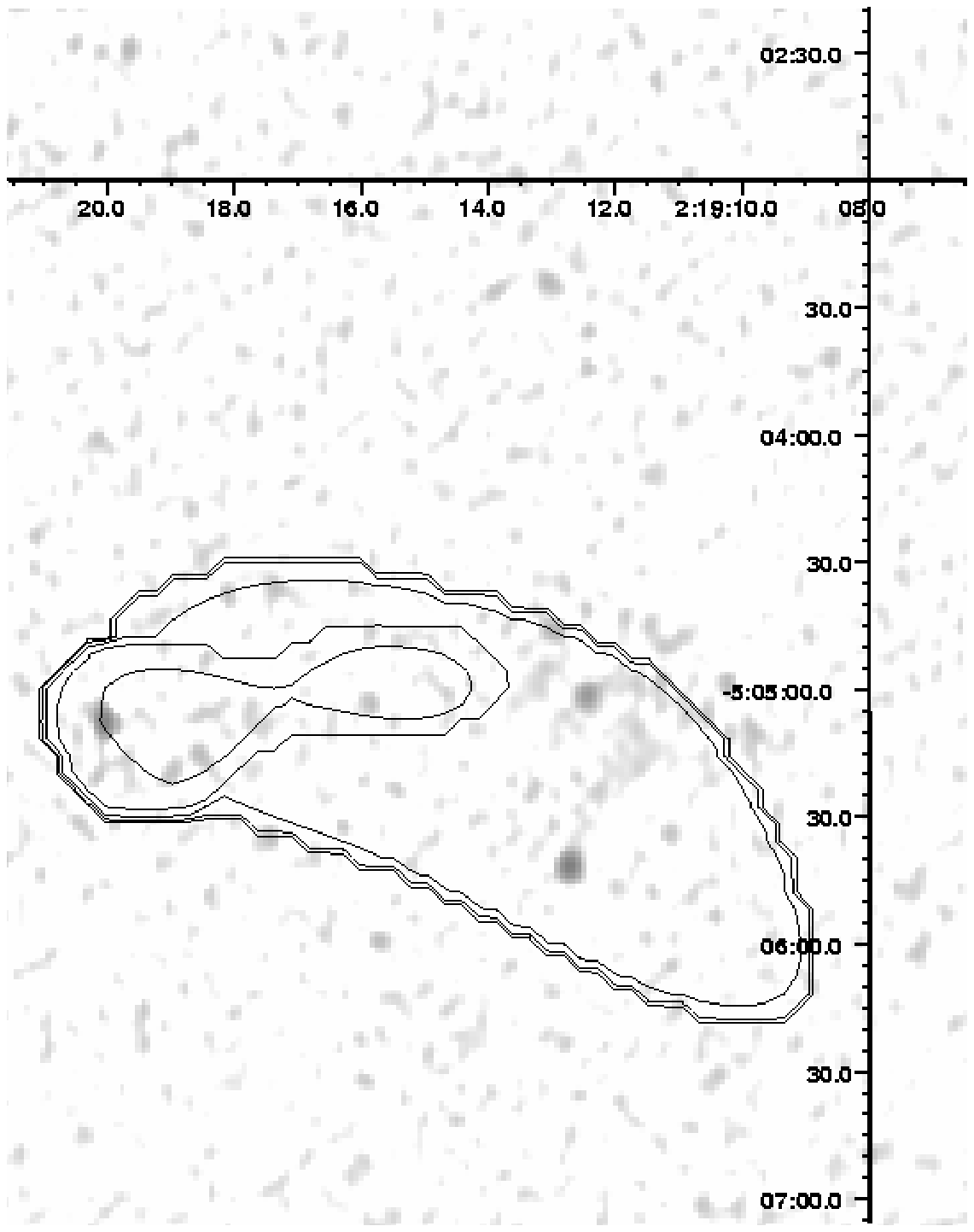} }
  \caption{Continued...  From upper left to lower right the cluster IDs are
    39, 57, 69, 70, 84.}  \end{figure*} 

In order to carry out this study we use an expected match in the shape of
the emission between the X-ray and radio source. The requirement for a match
in the orientation between X-ray and radio elongation to within $10^\circ$
substantially reduces a chance correspondence and is therefore used here to
build the best-practice examples of such matches, which later can be used to
treat more complex cases.  We used the VLA survey of the field at 1.4 GHz to
identify the radio sources (Simpson et al. 2006). The RMS of the image is
12--20$\mu$J.  There has been 14 complex morphology radio sources detected
inside X-ray selected clusters in SXDF, all shown in Fig.\ref{f:radio}.
Therefore a positional and azimuthal match is subject to a chance alignment
of 0.7\%, which can therefore be rejected with high confidence. With this
method we have found three X-ray sources (IDs 25,28 and 70), which emission
is entirely matched to a radio source. Three additional sources match
substructure detected in the X-ray images (IDs 4, 39, 69) and the original
sources had to be split (introducing new sources 90, 91 and 92 assigned to
X-ray jets) to ensure unbiased flux estimates of both components. In case of
cluster ID=69 and X-ray jet ID=92 the identification revealed different
redshifts of the counterparts, therefore increasing the number of X-ray jets
detected without detecting the cluster emission to 4 objects. This source
has also been discussed in Geach et al. (2007) and Tu et al.  (2009). All
four objects also exhibit a match in spatial extent between radio and
X-rays, which supports a physical link. The properties of the X-ray emission
associated with radio jets are summarized in Tab.\ref{t:radio}. Column 1
lists object ID, columns 2 and 3 list the coordinates of center of X-ray
emission, column 4 reports the spectroscopic redshift of the radio galaxy,
column 5 reports the X-ray flux in units of $10^{-15}$ ergs cm$^{-2}$
s$^{-1}$ and column 6 displays the corresponding rest-frame luminosity in
units of $10^{42}$ ergs s$^{-1}$.  We used the power law model with photon
index equal 2 in deriving the flux estimates and calculating the
K-correction for X-rays and a $\alpha=0.7$ index for the radio. Col (7)
lists the radio counterpart in the catalog of Simpson et al. (2006), col (8)
reports the flux ($F_r$) at 1.4GHz in mJy, col (9) reports the rest-frame
radio luminosity ($L_r$) at 1.4GHz, calculated using the following formulae:
$L_r = 4 \pi D_L^2 F_r (1+z)^(\alpha-1)$. The luminosity distance ($D_L$) is
calculated using the redshift listed in col.  4. The radio sources
responsible for most of the IC emission are at the $10^{25}$ W/Hz level (the
effect is detected from 20\% of all such radio sources), which are
characterized by the volume abundance of $\sim10^{-6}$ Mpc$^{-3}$ dex$^{-1}$
at redshifts near 1.  Comparison to theoretical model of Celotti \& Fabian
(2004), the IC effect detected in our survey is produced by the abundant
sources at the faint end of the radio luminosity function they considered.
This implies that the predictions in their Fig.3 need to be rescaled on
X-axis by factor of 10.  We can directly compute the required factor using
our X-ray and radio measurements, as reported in col.(9). Indeed the
obtained ratio is larger than 1. We have already dropped the factor
associated with the assumption of the evolution of the magnetic field (which
lowers the inferred X-ray flux for a given radio flux), which adds another
factor of 4, so on average a factor of 10 stronger production of X-rays
compared to a conservative assumption in Celotti \& Fabian (2004) is
observed. All six radio sources listed in Tab.\ref{t:radio} are considered
in detail as a part of the 
sample of Vardoulaki et al. (2008). At the same frequency, there is a good
agreement with NVSS measurements and only one source (radio ID=7) has a
steep spectrum (or much larger flux at lower frequencies compared to our
estimate here). Thus, our conclusion on a factor of 10 stronger production
of X-ray is neither an artifact, nor a result of using a different frequency
band compared to Celotti \& Fabian (2004).

\begin{figure}
  \includegraphics[width=8.cm]{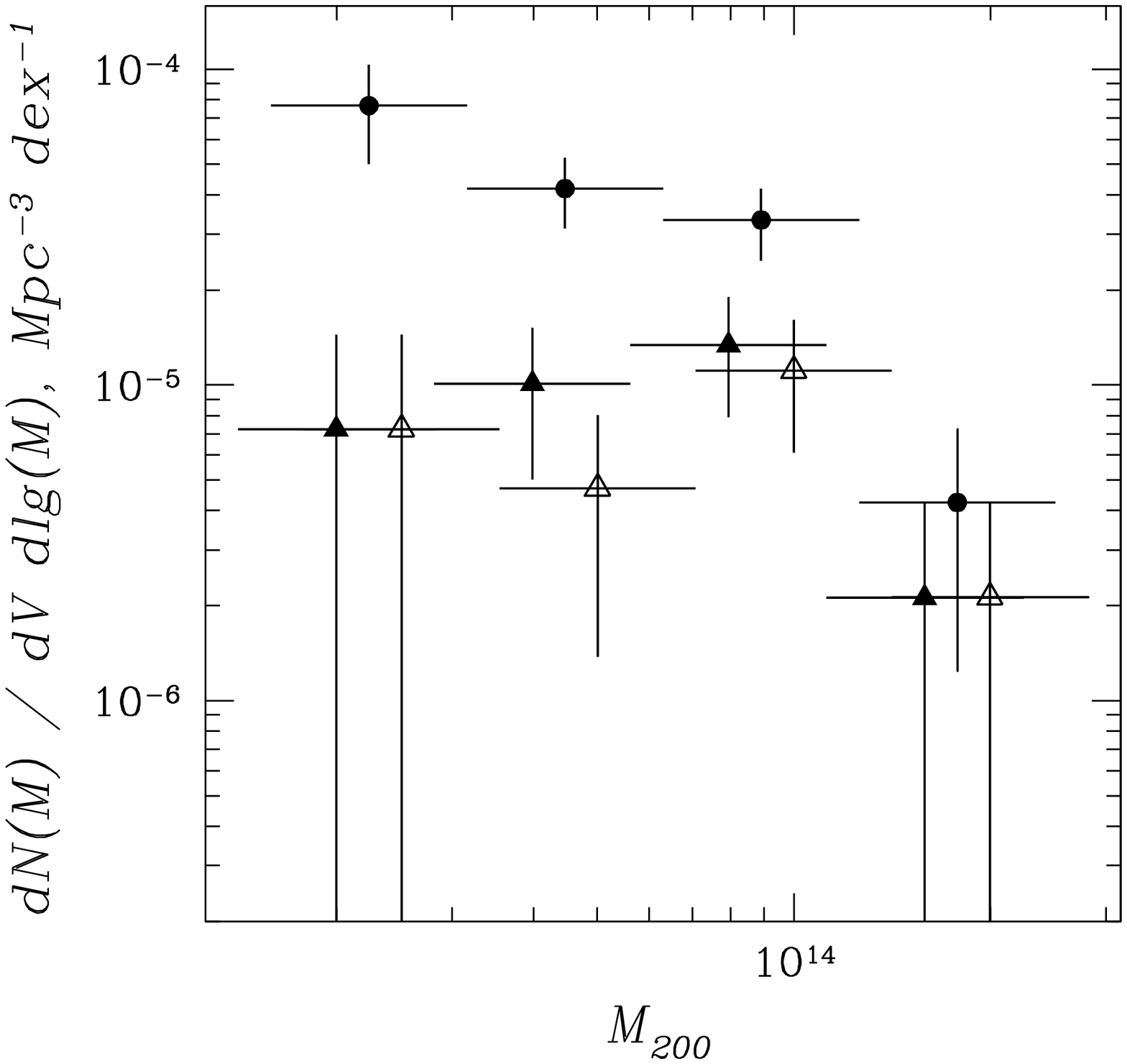} \caption{Mass
    function of $z<1$ identified X-ray emitting halos (solid circles with
    error bars) and those selected to have a radio galaxy from a
    luminosity-limited sample. Solid triangles with error bars show halos
    with the spectroscopically confirmed radio galaxies within $r_{200}$ and
    open triangles with error bars show matching within $0.2 r_{200}$, but
    keeping the galaxies with consistent photometric redshift estimate but
    having no spectroscopic information.  \label{hod}}
\end{figure}

One of the X-ray jets is remarkably bright in X-rays (ID 25). It would be
the most luminous cluster in the field, while an optical counterpart is
barely detected. The X-ray jet is located near an X-ray detected group into
which its host is probably accreting. The X-ray luminosity of the dominant
group is an order of magnitude fainter than that of the X-ray jet.  In
Fig.\ref{f:radio} we present all 14 complex morphology radio sources located
inside the extended X-ray emission.

In comparing the radio galaxy catalog to the catalog of X-ray clusters we
note that a number of these sources match and the chances for the X-ray
cluster to host a radio galaxy seem to increase with cluster mass.  In order
to characterize that we used our modeling of the survey to compute both the
mass function of the full sample and the mass function of X-ray clusters
that host a radio galaxy. We have excluded the 4 cases where X-ray emission
is caused by IC. In calculating the mass function, we take into account the
volume of the survey towards each cluster mass. For radio galaxies, we
select the luminosity-limited sample of $L_{1.4 GHz}>5\times10^{23}$ W/Hz,
which is valid to z of 1, given our flux limit of 100$\mu$Jy. We therefore
restricted the cluster selection and volume computation to a redshift of 1.
The limiting luminosity is located in the radio source population dominated
by FR I's (core brightened), which therefore justifies our use of limits for
a point source. Finally, since spectroscopic identification of radio catalog
is not complete, we calculate two examples of matching: one is by taking the
spectroscopically identified radio galaxies with redshifts matching that of
the cluster and the other is by assuming that once the radio galaxy is
located within the $0.2r_{200}$ it belongs to the cluster. There 15 galaxies
in total that fulfil this criterium, and after applying the radio luminosity
threshold there is 9 left, which we used to compute the marked mass
function. Only one system has a photometric redshift for the radio galaxy,
but the galaxy is also on the red-sequence, suggesting that the association
is real. The results are quantitatively similar and demonstrate in
Fig.\ref{hod} that indeed the probability to observe a radio galaxy
increases with mass of the halo. These results of a direct HOD determination
for radio AGNs are in good agreement with clustering analysis of 2SLAQ LRG
survey (Wake et al.  2008), performed at redshifts near 0.6 and a similar
selection of radio power.

\subsection{Cluster counts} 

It is common to characterize a cluster survey by its area as a function of
the limiting flux (which we do in Fig.\ref{f:area}) and present the results
as a relation between a cumulative surface density of clusters above a given
flux limit vs the flux value, the cluster $\log(N>S)-\log(S)$ (e.g.  Rosati
et al. 1998). The details of our calculation, which is shown in
Fig.\ref{f:logn}, are outlined in Finoguenov et al.  (2007). In addition to
the 57 identified sources, 9 sources were located in the area with
insufficient optical data due to either survey geometry or a presence of the
bright star.  In calculation of the upper limit on the $\log(N>S)-\log(S)$
we have added those sources using the typical flux extrapolation for our
apertures of 1.2. Sources identified as X-ray counterparts of radio jets
were not considered for $\log(N>S)-\log(S)$. 11 sources locate within the
area of best photometric data, but still lacking identification, were not
considered in the $\log(N>S)-\log(S)$. The computed uncertainties in
$\log(N>S)-\log(S)$ are purely statistical.  The comparison of the
$\log(N>S)-\log(S)$ to COSMOS results of Finoguenov et al. (2007) reveals a
good agreement at low fluxes, while at fluxes exceeding $5\times10^{-15}$
there is a lack of sources in the SXDF, compared to most previous surveys.
The variation of a statistics of bright sources is driven by the sample
variance in such fields (Hu \& Kravtsov 2003) and is very important for the
field-to-field comparison (e.g.  McCracken et al. 2007) on how clustering
affects the conclusions regarding galaxy evolution.  We present the previous
modeling of the $\log(N>S)-\log(S)$ of Rosati et al. (2002), which describes
well the cluster counts above $10^{-14}$ ergs cm$^{-2}$ s$^{-1}$.  The
short-dashed line is result of combining the adopted scaling relations,
WMAP5 concordance cosmology (Komatsu et al. 2009) and a cosmological code of
Peacock (2007).  Since it clearly overpredicts the observed counts and
previously published cluster counts, we considered the effect of excluding
low-luminosity ($L_x<10^{42}$ ergs s$^{-1}$) or high-redshift clusters
($z>1.2$) or both.  None of these attempts were successful in providing a
satisfactory solution.  In order to match the observations, we adopted a 5\%
lower value of $\sigma_8$, with a corresponding model prediction shown as a
solid line.  The role of SXDF in implying a change in the cosmological
parameters, is however moderate, since the small size of the field causes
large deviations at the bright end of the $\log(N>S)-\log(S)$.

\subsection{Sample characteristics}

\begin{figure}
\includegraphics[width=8.cm]{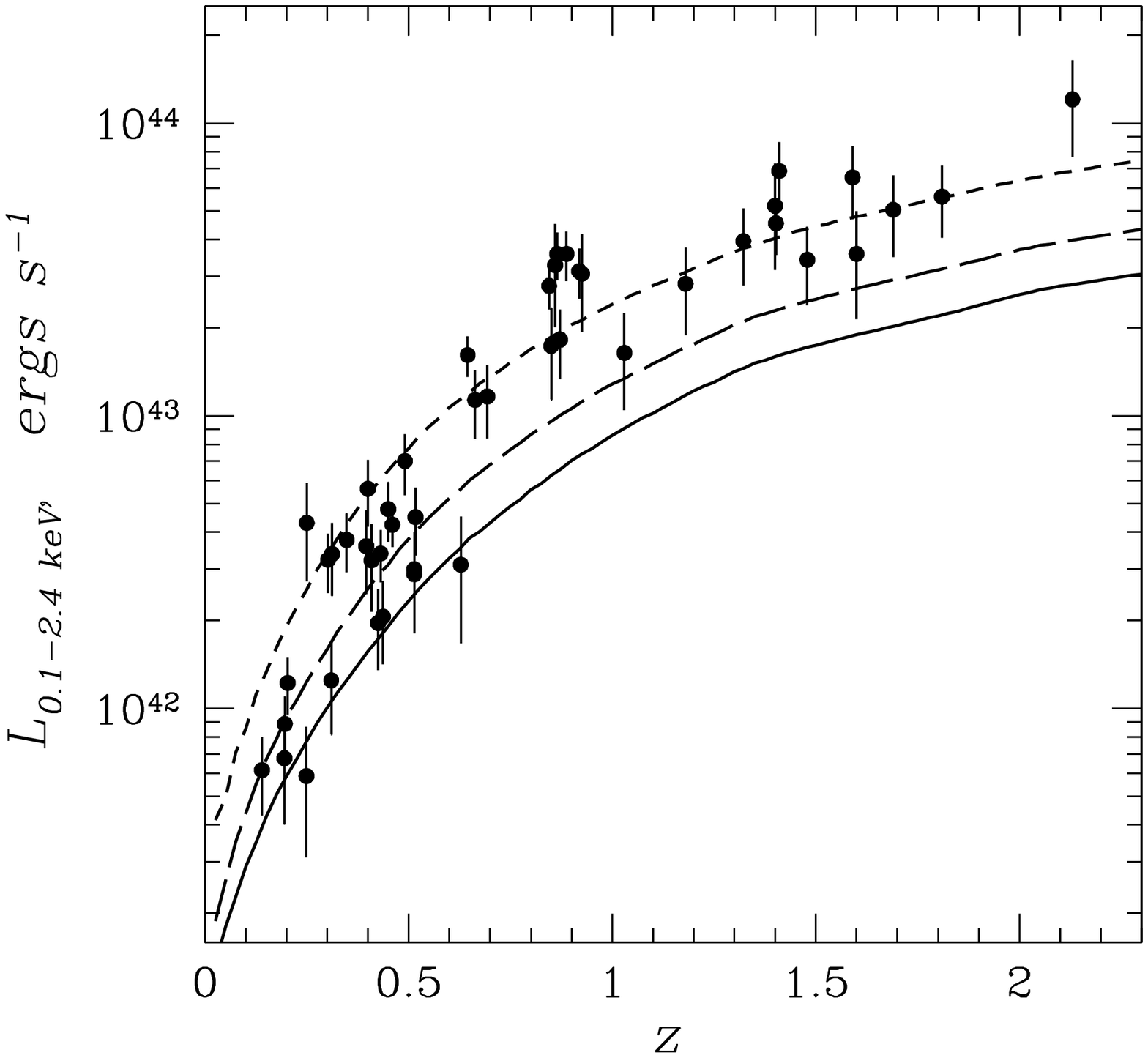}

\caption{Illustration of the cluster luminosity probed as a function of
cluster redshift in the SXDF. Filled circles represent the
detected clusters with error bars based on the statistical errors in the
flux measurements only. Short-dashed, long-dashed and solid black lines show
the flux detection limits of -14.5dex -14.8dex and -15.0dex associated with
90, 50 and 10\% of the total area, respectively.
\label{f:lxcat}}
\end{figure}

In Fig.~\ref{f:lxcat} we plot the observed characteristics of the SXDF
cluster sample together with detection limits implied by both survey depth
and our approach to search for clusters of galaxies. 

\begin{figure}
\includegraphics[width=8.cm]{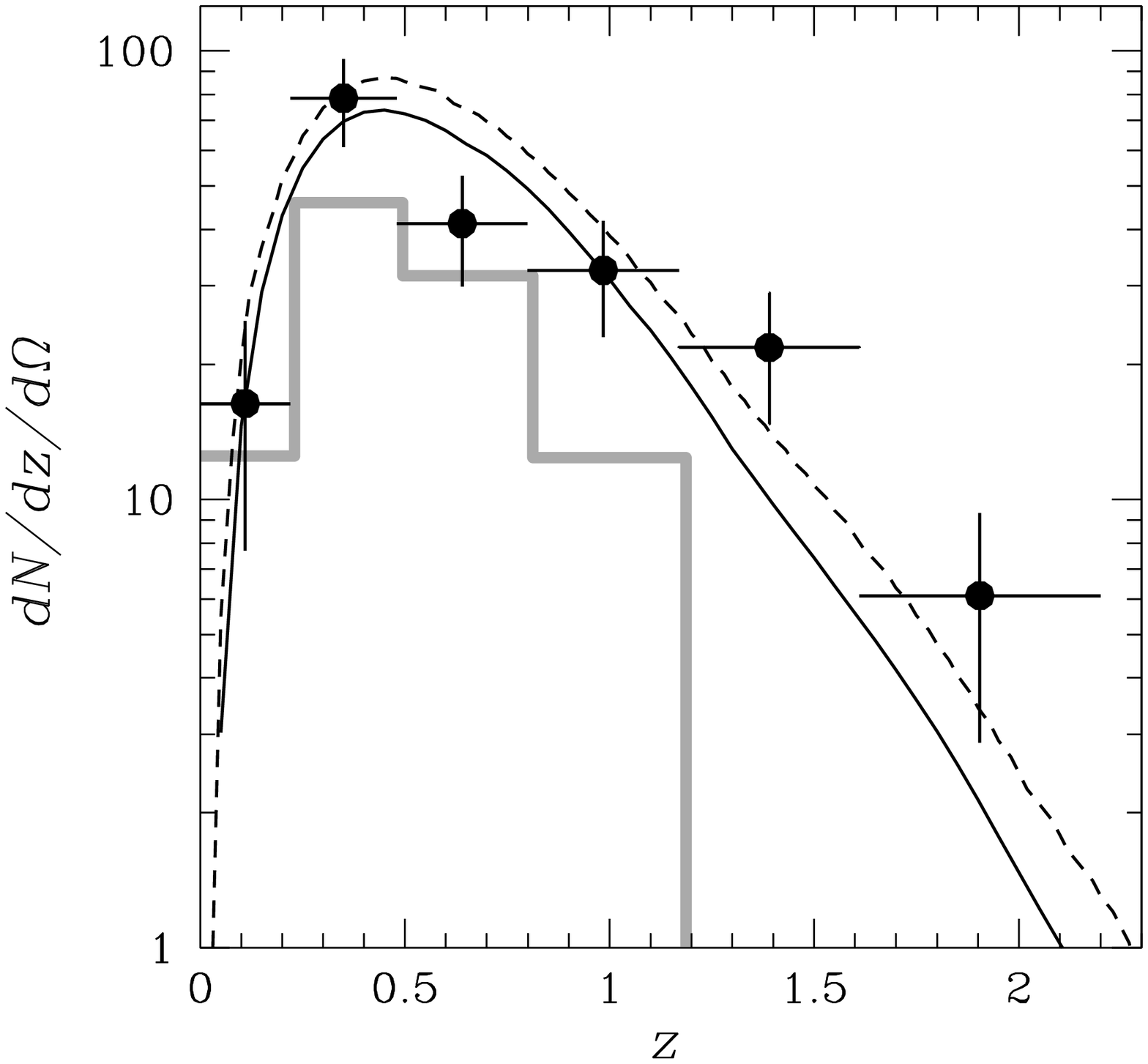}
\caption{Differential redshift distribution ($dN/dz/d\Omega$ per square
  degree) of 57 identified X-ray groups and clusters of galaxies in
  the SXDF (points with error bars). The thick grey histogram shows the
  spectroscopically confirmed systems. Short-dashed black curve shows the
  model prediction adopting WMAP5 cosmology. Solid line shows the WMAP5
  prediction with a reduced by 5\% value of $\sigma_8$. Similar to
  Fig.\ref{f:logn}, we are accounting for the incomplete identification by
  the 20\% upward correction for the data. 
  \label{f:dndz}}
\end{figure}

In Fig.~\ref{f:dndz} we report the redshift distribution of the identified
X-ray structures and attempt its modelling, assuming WMAP5 cosmology and
using the adopted scaling relations and their evolution. Compared to a
summary of scaling relations, presented in Finoguenov et al. (2007), we
adopted a direct $L_X-M$ calibrations of Rykoff et al. (2008), which match
well the results of COSMOS (Leauthaud et al. 2009). To compute the
K-correction, we still need $L-T$ relation, for which we adopt

\begin{equation}
  kT/keV=0.2+6\times10^{(lg(L_X/E_z/ergs/s)-44.45)/2.1)}
\end{equation}

and a fixed metalicity of 0.3 solar. The procedure for the flux
extrapolation is the same as in Finoguenov et al. (2007). The WMAP5
concordance model prediction for the dN/dz is shown in Fig.~\ref{f:dndz} as
a dashed curve and our best fit model to log(N)--log(S) data is shown as
solid curve. We show the results of our spectroscopic follow-up (grey
histogram) and account for incompleteness in our red sequence identification
by increasing the counts by a factor of 1.2, which accounts for the area
with lack of optical data. Removing the contribution of low luminosity
($L_X<10^{42}$ ergs s$^{-1}$) systems produces negligible result. The
largest deviation between the data and the model is the lack of clusters in
the 0.6--1.  redshift range, which can also be seen in Fig.\ref{f:lxcat}.
The follow-up of the candidates (summarized as a grey histogram in
Fig.~\ref{f:dndz}) is quite complete at those redshifts, so it might be an
effect of large-scale structure and shall be investigated further through a
comparison to other surveys like COSMOS.  The number of missing clusters at
those redshifts is around 10 similar to the deficit on $\log(N>S)-\log(S)$
at high fluxes. The number of our high-z candidates is consistent with the
cosmological expectation.

\subsection{X-ray luminosity function}\label{xfun}

The procedure of calculating the luminosity function is similar to COSMOS
(Finoguenov et al. 2007).

\begin{figure}
\includegraphics[width=8.cm]{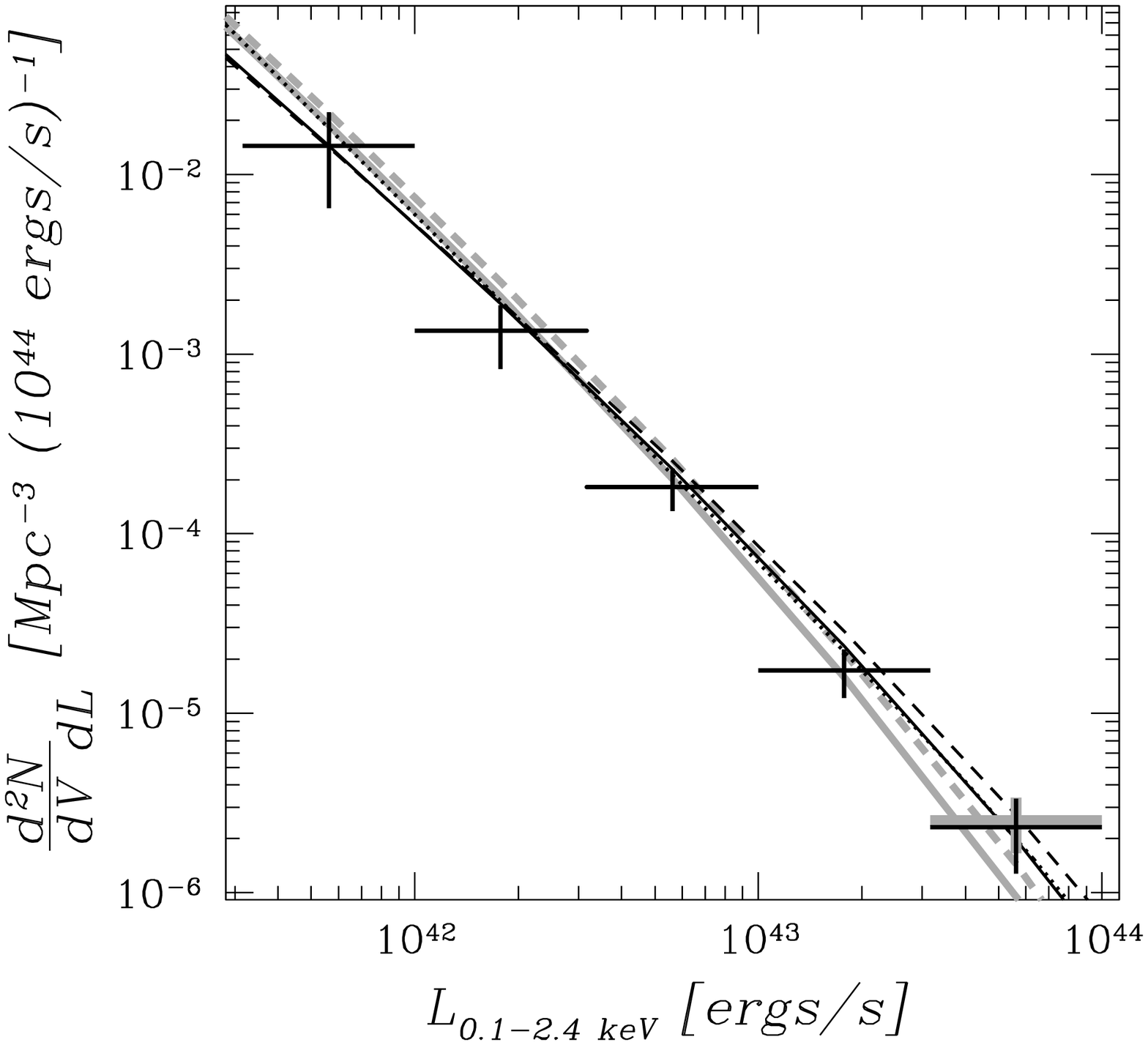}

\caption{Luminosity function of clusters in the SXDF field. Black
crosses indicate the data in the redshift range 0.2--1.0 and grey point
shows the data in the redshift range 1--2.5, which is the first measurement
reported for $z>1$. We apply the 20\% upward correction for incompleteness
of cluster identification of the field.
The dotted line shows the best fit to the
COSMOS data over $0<z<1.2$ (Finoguenov et al. 2007). The dashed lines
shows the WMAP5 
prediction for the luminosity function in the 0--1 (black) and 1--2.5 (grey)
redshift range. Solid lines shows the change in the model due to a 5\%
reduction  in $\sigma_8$ value.
\label{f:lxfunc}}
\end{figure}

In Fig.~\ref{f:lxfunc} we present the luminosity function of SXDF clusters
in the $0.2<z<1$ and $1<z<2.5$ redshift range. The choice of low redshift of
0.2 is due to incompleteness of the follow-up at low redshifts and we also
make an upward correction of the data by 1.2 accounting for incomplete
coverage of the X-ray data by the optical data. The SXDF results for the
$0.2<z<1$ compares well to the COSMOS results in the $0<z<1.2$ range, which
are also shown in the figure (dashed line). In Fig.~\ref{f:lxfunc} we also
show the prediction of our cosmological modelling and the assumed scaling
relations. The model describes well both the luminosity function, and its
redshift evolution.

We have tested the effects of log-normal scatter on the luminosity function
with $\delta lg(L_X)=0.2$ (Vikhlinin et al. 2009) and found them to be
important only at $L_x>10^{44}$ ergs s$^{-1}$. 

The sensitivity towards an assumption of $\sigma_8$ value is not so large
for low-mass systems, and in consistency with previous tests, we show the
prediction of a 5\% reduced value of $\sigma_8$ in Fig.~\ref{f:lxfunc}
(solid line). So, it is difficult to see if a particular part of XLF is
causing a requirement for lowering the $\sigma_8$ value. Changing $\Omega_m$
value would require a self-consistent recalibration of data, which is beyond
the scopes of this work, while the current dataset is sensitive only to
changes in $\Omega_m$ exceeding 10\%. Using galaxy groups one can break the
degeneracy between $\Omega_M$ and $\sigma_8$ present in fitting the cluster
counts. The biggest remaining uncertainty is scatter in the scaling
relations for galaxy groups, which is not well known.

%

\section{Conclusions}\label{resume}

We have searched for extended X-ray emission in the SXDF and presented the
catalog of identified X-ray groups and clusters of galaxies. Our analysis of
the extended X-ray emission in the Subaru-XMM Deep Field revealed a new
class of sources, associated with the inverse Compton emission from radio
lobes.  For extended objects identified as galaxy clusters, we derive the
statistical properties of the survey and compare them to published results
on COSMOS (Finoguenov et al. 2007) and the prediction of current best-fit
cosmology and cluster scaling relations.  We considered $\log(N>S)-\log(S)$,
dn/dz and XLF tests. $\log(N>S)-\log(S)$ test showed that SXDF lack extended
X-ray sources brighter than $5\times 10^{-15}$ ergs cm$^{-2}$ s$^{-1}$,
compared to other surveys, which we attribute to sample variance. XLF is in
good agreement with COSMOS and is well modelled, but somewhat more uncertain
due to incompleteness of the identification in SXDF field. Comparing the
WMAP5 cosmology together with the scaling relations for clusters to the
cluster counts, we detect a sensitivity of the sample towards present
uncertainty in the cosmological parameters and illustrate it by changing the
value of $\sigma_8$ by 5\% to provide a best fit to our data and showing the
change in the prediction for both $\log(N>S)-\log(S)$, dn/dz and XLF tests.

\section{Acknowledgments}
In Germany, the XMM--Newton project is supported by the Bundesministerium
fuer Wirtschaft und Technologie/Deutsches Zentrum fuer Luft- und Raumfahrt
(BMWI/DLR, FKZ 50 OX 0001). Part of this work was supported by the Deutsches
Zentrum f\"ur Luft-- und Raumfahrt, DLR project numbers 50 OR 0207 and 50 OR
0405.  AF acknowledges support from Spitzer UDS Legacy program to UMBC. AF
thanks Andy Fabian for useful discussions regarding the X-ray jets. AF
thanks the University of Leicester for the hospitality during his frequent
visits. The authors thank the referee for useful comments on the manuscript.

\label{lastpage}

\include{cat_tab3_astroph}

\include{zglx_astroph}

\include{xjets}

\end{document}

%% file: cat_tab3_astroph.tex
\begin{deluxetable}{lccccrrcccccccc}
\rotate
\tablewidth{0pt}
\footnotesize
\tablecaption{Catalog of the SXDF X-ray selected galaxy groups.\label{t:ol}}
\tablehead{
\colhead{ } &
\colhead{R.A} &  
\colhead{Decl.} & 
\colhead{z} &
\colhead{flux  $10^{-14}$} &
\colhead{L$_{\rm 0.1-2.4 keV}$} & 
\colhead{M$_{200}$} & 
\colhead{$r_{200}$} & 
\colhead{ }& 
\colhead{ }& 
\colhead{flux} &
\multicolumn{2}{c}{red sequence} &
\colhead{median} &
\colhead{Ueda }\\
\colhead{ID} & 
\multicolumn{2}{c}{Eq.2000} & 
\colhead{ } & 
\colhead{ergs cm$^{-2}$ s$^{-1}$} & 
\colhead{$10^{42}$ ergs s$^{-1}$} & 
\colhead{$10^{13}$ M$_\odot$} & 
\colhead{$\prime$} &
\colhead{flag} &
\colhead{N(z)} &
\multicolumn{2}{c}{significance} &
\colhead{redshift} &
\colhead{photo-z} &
\colhead{ ID}
\\
\colhead{(1)}&
\colhead{(2)}&
\colhead{(3)}&
\colhead{(4)}&
\colhead{(5)}&
\colhead{(6)}&
\colhead{(7)}&
\colhead{(8)}&
\colhead{(9)}&
\colhead{(10)}&
\colhead{(11)}&
\colhead{(12)}&
\colhead{(13)}&
\colhead{(14)}&
\colhead{(15)}
}
\startdata
SXDF01XGG & 34.68284 & -5.54973 & 0.378& 1.11$\pm$0.19 & 8.95$\pm$1.54 & 7.30$\pm$0.73 & 2.5 & 2 & 6&	 5.8 &   7.6 &  $0.29^{+0.05}_{-0.05}$ & 0.40& 0876 0889\\
SXDF03XGG & 34.33485 & -5.48511 & 0.382& 1.32$\pm$0.33 & 10.94$\pm$2.72 & 8.21$\pm$1.17 & 2.5 & 1 & 5&	 4.0 &   2.3 &  $0.22^{+0.03}_{-0.06}$ & 0.35\\
SXDF04XGG & 34.47560 & -5.45160 & 0.693& 0.31$\pm$0.09 & 11.69$\pm$3.30 & 6.55$\pm$1.05 & 1.5 & 2 & 3&	 3.4 &   8.7 &  $0.61^{+0.05}_{-0.06}$ & 0.70 & 0621\\
SXDF06XGG & 34.73260 & -5.47054 & 0.451& 0.52$\pm$0.12 & 6.52$\pm$1.51 & 5.68$\pm$0.76 & 2.0 & 1 & 4&	 4.3 &   8.5 &  $0.35^{+0.06}_{-0.03}$ & 0.45& 0934\\
SXDF07XGG & 34.60464 & -5.41436 & 0.646& 0.84$\pm$0.13 & 24.67$\pm$3.82 & 10.68$\pm$0.96 & 1.9 & 1 & 6&	 6.5 &   8.4 &  $0.56^{+0.04}_{-0.07}$ & 0.65 & 0784\\
SXDF08XGG & 34.36312 & -5.41925 & 0.645& 0.51$\pm$0.08 & 15.49$\pm$2.48 & 8.09$\pm$0.75 & 1.7 & 2 & 3&	 6.4 &   5.3 &  $0.53^{+0.05}_{-0.04}$ & 0.65\\
SXDF10XGG & 34.19894 & -4.55896 & 0.409& 0.32$\pm$0.11 & 3.21$\pm$1.07 & 3.84$\pm$0.72 & 1.9 & 3 & 2&	 2.9 &   1.5 &  $0.30^{+0.03}_{-0.01}$ & 0.40\\
SXDF14XGG & 34.49784 & -4.63107 & 0.396& 0.42$\pm$0.13 & 3.88$\pm$1.22 & 4.35$\pm$0.78 & 2.0 & 5 & 2&	 3.2 &   3.3 &  $0.30^{+0.03}_{-0.10}$ & 0.35\\
SXDF15XGG & 34.23301 & -4.65666 & 0.437& 0.18$\pm$0.05 & 2.06$\pm$0.64 & 2.88$\pm$0.51 & 1.6 & 1 & 1&	 3.6 &   4.7 &  $0.34^{+0.06}_{-0.12}$ & 0.40\\
SXDF16XGG & 34.35674 & -4.65945 & 0.196& 0.55$\pm$0.13 & 0.92$\pm$0.22 & 2.15$\pm$0.30 & 2.8 & 1 & 3&	 4.2 &   1.9 &  $0.15^{+0.02}_{-0.02}$ & 0.25 & 0453 \\
SXDF18XGG & 34.53927 & -4.67350 & 0.312& 0.69$\pm$0.19 & 3.39$\pm$0.94 & 4.30$\pm$0.68 & 2.4 & 2 & 2&	 3.6 &   2.8 &  $0.27^{+0.02}_{-0.04}$ & 0.30\\
\multicolumn{15}{c}{The rest of the table will be released after acceptance of the paper} \\
\enddata
\\
$^a$ -- stellar halo
\end{deluxetable}

%% file: zglx_astroph.tex
\begin{deluxetable}{cccc}
\footnotesize
\tablecaption{Spectroscopic redshifts of cluster member galaxies.\label{t:zgal}}
\tablewidth{0pt}
\tablehead{
\colhead{R.A} &  
\colhead{Decl.} & 
\colhead{z} &
\colhead{cluster}\\
\multicolumn{2}{c}{Eq.2000} & 
\colhead{ } & 
\colhead{ID} \\
\colhead{(1)}&
\colhead{(2)}&
\colhead{(3)}&
\colhead{(4)}
}
\startdata
 34.672458 & -5.534833 & 0.381 & 1\\
 34.675250 & -5.547528 & 0.381 & 1\\
 34.678875 & -5.580494 & 0.375 & 1\\
 34.681125 & -5.553822 & 0.375 & 1\\
 34.690000 & -5.545900 & 0.375 & 1\\
 34.690667 & -5.548917 & 0.383 & 1\\
 34.315625 & -5.488881 & 0.382 & 3\\
 34.328500 & -5.491192 & 0.381 & 3\\
 34.332292 & -5.502233 & 0.383 & 3\\
 34.333792 & -5.492828 & 0.381 & 3\\
 34.359750 & -5.507003 & 0.383 & 3\\
 34.488667 & -5.465550 & 0.694 & 4\\
 34.470042 & -5.438756 & 0.695 & 4\\
 34.475771 & -5.451581 & 0.690 & 4\\
 34.735500 & -5.472017 & 0.452 & 6\\
 34.737667 & -5.472025 & 0.450 & 6\\
 34.745083 & -5.497194 & 0.453 & 6\\
 34.758708 & -5.467483 & 0.450 & 6\\
 34.579429 & -5.396031 & 0.646 & 7\\
 34.588342 & -5.421225 & 0.643 & 7\\
 34.598000 & -5.416903 & 0.647 & 7\\
 34.604929 & -5.421614 & 0.646 & 7\\
 34.607771 & -5.422281 & 0.647 & 7\\
 34.612500 & -5.418975 & 0.646 & 7\\
 34.346642 & -5.405492 & 0.644 & 8\\
 34.346871 & -5.414247 & 0.643 & 8\\
 34.375917 & -5.425039 & 0.648 & 8\\
 34.202500 & -4.555364 & 0.409 & 10\\
 34.204000 & -4.555958 & 0.409 & 10\\
 34.489558 & -4.650786 & 0.394 & 14\\
 34.513667 & -4.625025 & 0.397 & 14\\
 34.239917 & -4.664153 & 0.437 & 15\\
 34.352625 & -4.662078 & 0.197 & 16\\
 34.359542 & -4.659900 & 0.195 & 16\\
 34.361042 & -4.662269 & 0.196 & 16\\
 34.531042 & -4.682667 & 0.314 & 18\\
 34.544083 & -4.710425 & 0.310 & 18\\
Abridged & & & \\
\enddata

\end{deluxetable}

%% file: xjets.tex
\begin{deluxetable}{cccccccccc}
\footnotesize
\tablecaption{Properties of extended X-ray emission associated with radio jets.\label{t:radio}}
\tablewidth{0pt}
\tablehead{
\colhead{ ID host } &
\colhead{R.A} &  
\colhead{Decl.} & 
\colhead{z} &
\colhead{flux  $10^{-15}$} &
\colhead{L$_{0.5-2 keV}$} &
\colhead{Simpson}&
\colhead{Flux}&
\colhead{$L_{1.4GHz}$ }&
\colhead{ $L_x \nu_x / L_r \nu_r$}\\
\colhead{cluster} & 
\multicolumn{2}{c}{Eq.2000} & 
\colhead{ } & 
\colhead{ergs cm$^{-2}$ s$^{-1}$} & 
\colhead{$10^{42}$ ergs  s$^{-1}$} &
\colhead{ ID}&
\colhead{ 1.4GHz, mJy}&
\colhead{ $10^{23}$ W/Hz}&
\colhead{$\times(1+z)^{-3.7}$}\\
\colhead{(1)}&
\colhead{(2)}&
\colhead{(3)}&
\colhead{(4)}&
\colhead{(5)}&
\colhead{(6)}&
\colhead{(7)}&
\colhead{(8)}&
\colhead{(9)}&
\colhead{(10)}
}
\startdata
SXDF90XGG & 34.47560 & -5.45160 & 0.693 &  1.66 $\pm$ 0.50  & 3.6$\pm$1.1 & 19 & 4.83 & $87\pm1$ & 4.2\\
SXDF25XGG & 34.24682 & -4.82231 & 1.179 &  6.85 $\pm$ 0.91  & 54.$\pm$7.1 & 7 & 9.6   & $605\pm2$ & 3.6 \\
SXDF28XGG & 34.49991 & -4.82799 & 0.192 &  1.24 $\pm$ 0.37  & 0.14$\pm$0.04 & 20 & 4.6 & $4.4\pm0.02$ & 12.\\
SXDF91XGG & 34.14713 & -4.91255 & 0.865 &  0.74 $\pm$ 0.27  & 2.8$\pm$1.0 & 12  & 6.59 & $200\pm1$ & 1.0$^a$\\
SXDF92XGG & 34.39369 & -5.22180 & 0.645 &  0.80 $\pm$ 0.23  & 1.5$\pm$0.4 & 33 & 2.37  & $36\pm0.4$ & 4.7 \\
SXDF70XGG & 34.35118 & -5.21950 & 0.919 &  1.89 $\pm$ 0.34  & 8.2$\pm$1.5 & 18 & 4.84  & $170\pm1$ & 3.1\\
\enddata
\\
$^a$ -- only one radio lobe is used in the X-ray estimate
\end{deluxetable}